\documentclass[a4,11pt]{iopart}

\usepackage{amstext}
\usepackage{amssymb}
\usepackage{iopams}
\usepackage{float}
\usepackage{epsfig}

\newcommand{\eps}{\epsilon}

\renewcommand{\Re}{{\rm Re}}
\usepackage{graphicx,xcolor}
\usepackage[latin1]{inputenc}
\usepackage[T1]{fontenc}
\usepackage[english]{babel}
\usepackage{ae,aecompl}

\newcommand{\smallmatrix}[1]{
  \text{\begin{footnotesize}
      $ \pmatrix{ %                                                                                                                           
        #1} $
\end{footnotesize}}}

\newcommand{\eqref}[1]{(\ref{#1})}

\usepackage[pdftex]{hyperref}
\usepackage{xcolor}
\begin{document}

\title[Multilane driven diffusive systems] {Multilane driven diffusive
  systems}

\author{A. I. Curatolo$^1$, M. R. Evans$^2$, Y. Kafri$^3$,  J. Tailleur$^1$}

\address{$^1$ Universit\'e Paris Diderot, Sorbonne Paris Cit\'e, MSC, UMR 7057 CNRS, 75205 Paris, France}

\address{$^2$ SUPA, School of Physics and Astronomy, University of
 Edinburgh,
Peter Guthrie Tait Road,
Edinburgh
EH9 3FD, Scotland}

\address{$^3$ Department of Physics, Technion, Haifa, 32000, Israel}

\eads{
  \mailto{agnese.curatolo@univ-paris-diderot.fr}
  \mailto{martin@ph.ed.ac.uk}, 
\mailto{kafri@physics.technion.ac.il},
\mailto{Julien.Tailleur@univ-paris-diderot.fr}}

  \date{today}
  \begin{abstract}
    We consider networks made of parallel lanes along which particles
    hop according to driven diffusive dynamics. The particles also hop
    transversely from lane to lane, hence indirectly coupling their
    longitudinal dynamics. We present a general method for
    constructing the phase diagram of these systems which reveals
    that in many cases their physics reduce to that of single-lane
    systems. The reduction to an effective single-lane description
    legitimizes, for instance, the use of a single TASEP to model the
    hopping of molecular motors along the many tracks of a single
    microtubule. Then, we show how, in quasi-2D settings, new
    phenomena emerge due to the presence of non-zero transverse
    currents, leading, for instance, to strong `shear localisation'
    along the network.
  \end{abstract}
  
  \maketitle
  
%  \tableofcontents
  
  \section{Introduction}

  Driven diffusive systems encompass a broad class of statistical
  physics models involving many interacting particles moving
  stochastically under dynamics that do not respect detailed
  balance~\cite{SZ,Mukamel00}. As such, the stationary states attained
  in these systems are {\em nonequilibrium stationary states} (NESS)
  and exhibit currents both at the macroscopic level (particle
  currents) and microscopic level (probability currents in
  configuration space).

One-dimensional realisations have proven particularly informative. The
general scenario is a one-dimensional lattice connected to particle
reservoirs of fixed density at each end.  The macroscopic current that
flows through the system is determined by an interplay between the
boundary reservoir densities and the bulk dynamics of the
particles. As pointed out by Krug \cite{Krug} this can lead to boundary-induced phase transitions,
wherein the macroscopic current can be controlled by the boundary
rather than bulk dynamics---such transitions have no counterpart in equilibrium
stationary states which contain no currents.

The first model of this kind to be studied in detail was the Totally
Asymmetric Simple Exclusion Process with open boundaries (TASEP) for which an
exact phase diagram was derived \cite{DDM92,DEHP93,SD93}.  For more
general models, however, exact solutions are difficult to come by,
therefore to make progress it is important to develop approximations
such as mean field theory \cite{DDM92} and heuristic approaches. For
one-dimensional models a particularly useful approach uses an Extremal
Current Principle (ECP) \cite{Krug,PS,Hager}. The principle (see
below) uses the relation $J(\rho)$, between the particle current $J$
and local particle density $\rho$, along with the boundary conditions
to derive the phase diagram. Its predictions are borne out by exact
phase diagrams, for example in the case of the TASEP.

It is noteworthy that the TASEP had first been introduced some years
ago as a model for ribosome dynamics in RNA translation
\cite{MacDon}. Since then variants of the basic model have been used
as a description of various biophysical transport process such as
molecular motors moving on microtubules \cite{AMP99,KL03}, fungal
hyphae growth dynamics \cite{SEPR07}, extraction of membrane tubes
\cite{Tailleur2009PRL} and the model has become a generic starting
point to describe transport processes \cite{Schad,traffic}.  However,
many of these processes require a more complicated geometry than a
single one-dimensional lattice.  For example, motorway traffic
involves several lanes with interchange between lanes and the motion
of motor proteins along cytoskeletal filaments often has multiple
tracks which may involve different hopping and boundary rates and
possibly motors moving in different directions \cite{GCAR12,NOSC05}.
Pedestrian flows often involve two ``multi-lane streets'' crossing
perpendicularly and a collective dynamics arises in the crossing
region \cite{Hilhorst}.  Inspired by these contexts  multilane
exclusion processes  have been considered by several authors~\cite{PS03,PS04,PK04,HS05,SKZ05,Juhasz07,RFF07,JHWW08,CGW08,JNHWW09,SARS10,YSM12,WJWW14}
(see for example \cite{EKST11} for an overview)
%\cite{}.  In particular
%systems consisting of two coupled TASEPs have been the subject of several
%studies~\cite{}.  More general
%multilane systems have also been studied~\cite{} and the
%hydrodynamics of coupled two-species systems has been considered
%in~\cite{.
and moreover exclusion processes on more complicated graphs such as
trees and networks have been formulated and results obtained
\cite{BM10,NKP11,MWE13}.  However, for these more complicated
geometries exact solutions are few and far between, therefore
approximate or heuristic approaches are essential.

In \cite{EKST11} a general class of two-lane models has been
considered under the main restrictions that: (i) The transport within
a lane is local and depends only on the densities within that lane and
not on those of neighbouring lanes.  (ii) The average transverse flux
of particles from one lane to the other increases with the density of
the departing lane and decreases with that of the arriving one
\cite{EKST11}.  Under these conditions it was shown, through a linear
stability analysis of the continuum mean-field equations for the
densities, how the phase diagram can be constructed.  This
construction turns out to be equivalent to an ECP which holds for the
total particle current $J_{\rm tot}$, defined as the sum of the
particle currents in each lane.

In this work we generalize this stability analysis to the multi-lane
case which naturally connects one-dimensional lattice gases to their
two dimensional counterparts or to more general network topolgies. In
addition to the \textit{longitudinal} currents flowing along each
lane, new steady \textit{transverse} currents flowing between the
lanes can be observed in such systems. Their impact on the
phenomenology of multi-lane system has not been studied so far. By
studying the eigenvalue spectrum of the linearised mean-field
equations for the densities, we show how an ECP holds for the total
longitudinal current. This result implies that a system with an
arbitrary number of lanes (such as the motion of molecular motors
along the protofilaments of microtubules) may effectively be described
by a one-lane system with generally a non-trivial current density
relation, validating an assumption that is commonly made, see for
example~\cite{CMZ11} and references therein. A consequence of this
effective description is that the transverse current does not enter
into the determination of the phase diagram. Nevertheless, transverse
currents can be associated with interesting new phenomena. For
instance, one can consider ``sheared'' systems, in which boundary
conditions impose different transverse currents at the two ends of the
system. For diffusive bulk dynamics, the transverse current varies
continuously, interpolating between its imposed boundary
conditions. As we show in this article, for systems driven in the
bulk, one may observe a discontinuity in the transverse current
corresponding to {\em shear localisation}.

The paper is organised as follows. In section 2, for sake of
completeness, we review the ECP for the single-lane TASEP. In section
3, we then present the general framework used in this article to
describe multi-lane systems: their description at the level of
continuum mean field equations.  In section 4 we present a linear
analysis of these equations and use these results to formulate a
general method for constructing phase diagrams, which amounts to a
generalised ECP. In section 5, we turn to concrete examples and
present numerical simulations of microscopic models. This allows us to
confirm our predictions as well as illustrate the new phenomenology
associated to steady transverse currents. Technical details and proofs
are left to the appendices.

\begin{figure}[H]
  \centering
  \includegraphics[width=.5\textwidth]{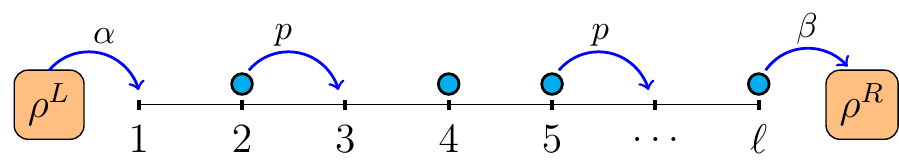}
  \caption{Illustration of the one-dimensional totally asymmetric exclusion process with open boundaries (TASEP)}
%  \caption{TASEP system}
  \label{fig:TASEP}
\end{figure}

\section{A brief recap of the TASEP}
\label{sec:recapTASEP}
Before turning to the full multilane problem it is useful to recall
the TASEP and its phase diagram. The TASEP consists of a
one-dimensional lattice of length $\ell$ with totally asymmetric
hopping dynamics of hard-core particles: at most one particle is
allowed at each site and particles can only move in the forward
direction with rate $p$ which we may take to be unity (see
fig.~\ref{fig:TASEP}).  At the left boundary particles enter with rate
$\alpha$ provided the first site is empty and when a particle arrives
at the right boundary it leaves with rate $\beta$.  For $\alpha, \beta
\leq 1$ these boundary conditions correspond to a left reservoir with
density $\rho^L =\alpha$ and a right reservoir with density $\rho^R
=1-\beta$. 
% The exact solution of the stationary state and phase diagram for
% this model has been worked out \cite{DDM92,DEHP93,SD93}. -> Already
% said in intro.

\begin{figure}[hbt]
  \centering
  \includegraphics{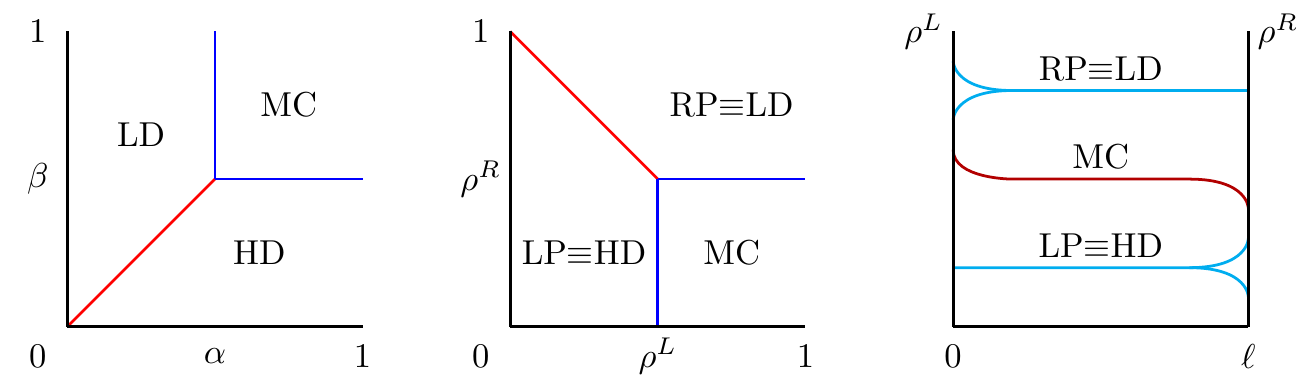}
  \caption{LEFT: Phase diagram of the TASEP where the control
    parameters are the injection rate $\alpha$ and ejection rate
    $\beta$ and the phases are Low Density (LD), High Density (HD) and
    Maximal Current (MC). CENTER: phase diagram where the control
    parameters are the left and right reservoir densities and the
    phases are Left Phase (LP), Right Phase (RP) and Maximal Current
    (MC). In both cases the blue lines represent a second order phase
    transition while the red lines represent a first order phase
    transition. RIGHT: profiles corresponding to left, right and
    maximal current phases.}
  \label{fig:PhDiag}
\end{figure}

The phase diagram for the TASEP in the large $\ell$ limit is
illustrated in Fig.~\ref{fig:PhDiag}.  The different phases that can
be observed are:
\begin{itemize}
\item a left phase (LP), corresponding to a bulk profile whose density is equal to that of the left reservoir and in which the current is
$J =\rho^L(1-\rho^L)$
\item a right phase (RP), corresponding to a  bulk profile whose density is equal to that of the right reservoir  and in which the current is
$J =\rho^R(1-\rho^R)$
\item a maximal current phase (MC), corresponding to a bulk profile
  whose density $\rho^M =1/2$ in which the current is maximised at
  $J=1/4$.
\end{itemize}
Note that historically the  Left and Right
phases have been referred to as  low density (LD) and high density (HD).
Here we use a different terminology that is more appropriate to systems that may include further phases.

To determine the bulk density of single lane systems, Krug~\cite{Krug}
introduced a maximal current principle which was later generalized to
an ECP~\cite{Hager} to describe systems in which the advected current
$J(\rho)$ has more than one extrema. In practice, this principle
states that, given two reservoir densities $\rho^L$ and $\rho^R$, the
dynamics select a constant plateau whose density $\rho^B$ is
intermediate between $\rho^L$ and $\rho^R$, and tends to maximize or
minimize the advected current $J(\rho^B)$:
\begin{equation}
  J(\rho^B)=\left\{
    \begin{array}{lr}
      \max_{\rho\in[\rho^R,\rho^L]} \ J(\rho) \qquad \text{if} \quad \rho^L>\rho^R \\
      \min_{\rho\in[\rho^L,\rho^R]} \ J(\rho) \qquad \text{if} \quad \rho^L<\rho^R\;. \\
    \end{array}
  \right.
  \label{extrcurr}
\end{equation}
For the TASEP, the current density relation is $J(\rho) =
\rho(1-\rho)$ which is most simply derived from a mean-field
consideration that a particle, present with probability $\rho$, moves
forward with rate 1 when there is an empty site ahead (which has
probability $1-\rho$). The current-density relation for the TASEP has
a single extremum which is a maximum of the current at $\rho=1/2$. The
ECP then allows one to easily derive the phase diagram shown in
Fig~\ref{fig:PhDiag}.  Note that for more general current density
relations, which may exhibit several extrema, the ECP predicts, in
addition to the Left, Right and Maximal Current Phases exhibited by
the TASEP, a new minimal current phase (mC). This phase corresponds to
a bulk profile whose density $\rho^m$ is a local minimum of the
current-density relation $J(\rho)$.

\begin{figure}[H]
  \centering
  \includegraphics{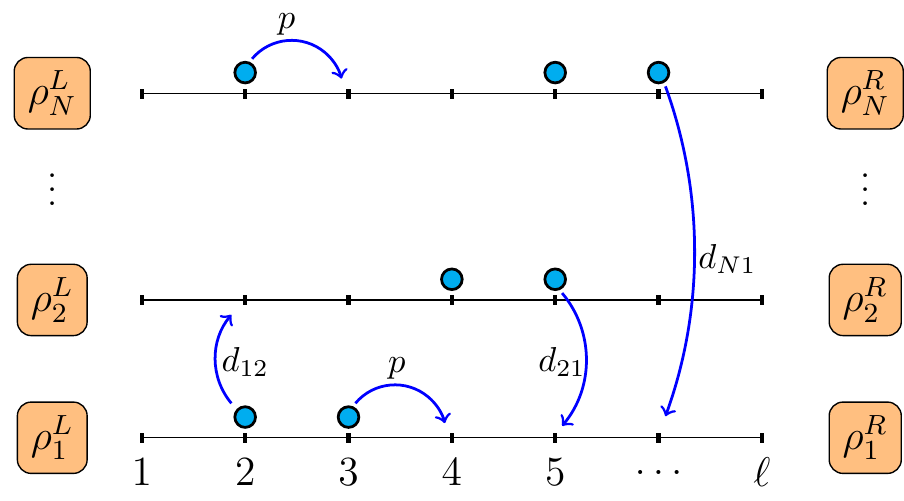}
  \caption{Schematic representation of $N$ parallel TASEPs which are
    connected at both ends to particle reservoirs of fixed
    densities. Particles hop along each lane, to the right, at rate
    $p$ and can hop from lane $i$ to lane $i\pm 1$ at rate $d_{i,
      i\pm1}$, giving rise to non-zero longitudinal and transverse
    currents.}
  \label{fig:Nlanes}
\end{figure}

\section{General framework}

\label{sec:analytics}

In this article, we consider driven diffusive systems in which
particles hop along or between $N$ parallel one-dimensional lattices
each of $\ell$ sites. At both ends, each of these $N$ `lanes' is in
contact with its own particle reservoir which acts as lattice site of fixed
mean density denoted by $\rho_i^{L/R}$ for the left/right boundary site of lane $i$. One simple system falling in this class consists of
the $N$ parallel TASEPs shown in Fig.~\ref{fig:Nlanes}.

As a starting point of our analysis, we assume that the dynamics of the
system is described by a set of $N$ coupled non-linear mean-field
equations for the density $\rho_i(x,t)$ of particles in lane $i$:
\begin{equation} \dot{\rho}_i(x,t)=-\partial_x[J_i(\rho_i(x,t))-D_i\partial_x\rho_i(x,t)]+\sum_{j\neq i}K_{ji}(\rho_j(x,t),\rho_i(x,t)),
  \label{mean-field}
\end{equation}
where $x=n/\ell$ is the position of site $n$ along each lane,
$J_i(\rho_i(x))$ and $-D_i\partial_x\rho_i$ are the advective and the
diffusive parts of the mean-field current along lane $i$, and $K_{ji}$
is the net transverse current flowing from lane $j$ to lane $i$ (with
$K_{ij}=-K_{ji}$). Throughout this article, we consider systems which
satisfy
\begin{equation}
  \partial_{\rho_i} K_{ij}(\rho_i,\rho_j)>0\quad\text{and}\quad   \partial_{\rho_j} K_{ij}(\rho_i,\rho_j)<0\;.
  \label{fluxes}
\end{equation}
Physically, this means that the net flux of particles from lane $i$ to lane $j$
increases with $\rho_i$ and decreases with $\rho_j$.

If the hopping rate of particles
along lane $i$ depends on the occupancies of lane $j\neq i$, then $J_i$
will depend on $\rho_j$ and not solely on $\rho_i$. In this article,
however, we restrict our attention to  the case where the hopping rate along one lane depends only on the occupancies of this lane, so that the
mean-field longitudinal current depends only on the density of this
lane: $J_i(\rho_i)$.

The derivation of (\ref{mean-field}) from the microscopic dynamics
models follows a standard mean-field approximation comprising
factorisation of all density correlation functions  which we review in
section~\ref{sec:numerics}.  For illustrative purposes we consider the
example presented in Fig.~\ref{fig:Nlanes} of coupled totally
asymmetric exclusion processes: particles hop forward longitudinally
along the lane $i$ from site $n$ to the next site $n+1$ if that site is
vacant with rate $p$; particles can hop transversely to a vacant site
at position $n$ in a neighbouring lane $i\pm 1$ with rate $d_{i, i\pm
  1}$.  Then $J_i(\rho_i) = p \rho_i(1-\rho_i)/\ell$, $D_i = p/(2
\ell^2)$ and $K_{i \pm 1, i} = d_{ i\pm 1\,, i}\rho_{i\pm 1}(1-\rho_i)
-d_{ i, i\pm 1}\rho_{i}(1-\rho_{i\pm 1})$.  The explicit form of
$J_i$, $D_i$ and $K_{ij}$ for other models is given in
section~\ref{sec:numerics}.

Given reservoir densities we want to
compute the average density profiles along each lane, i.e.  the
steady-state solutions of Eq.~(\ref{mean-field}) satisfying
$\rho_i(0)=\rho_i^L$ and $\rho_i(1)=\rho_i^R$. Solving this set of $N$
coupled non-linear equations is, however, very hard even in the steady
state.  Therefore we first consider the possible homogeneous solutions
to Eq.~(\ref{mean-field}) which will be valid in the bulk i.e. far
from the boundaries.  We will refer to these solutions as
\emph{equilibrated plateaux} since their densities are balanced by the
exchange of particles between lanes.  

As we shall see, these solutions, which we properly define in
section~\ref{sec:EPdef}, play a major role in constructing the form of
the steady-state profiles and the phase diagrams. We show in
section~\ref{sec:EqnonEqres} how stationary profiles are typically
made of equilibrated plateaux connected at their ends either to other
plateaux (forming shocks) or directly to the reservoirs, see
Fig.~\ref{fig:RLSphase}. For this to hold, the equilibrated plateaux
have to be dynamically stable which we show in
section~\ref{sec:dynstab} to be true for systems
satisfying  conditions~\eqref{fluxes}.

\subsection{Plateaux solutions}
\label{sec:EPdef} The plateaux solutions (homogeneous, steady-state solutions
denoted  $\rho_i(x)=\rho_i^{\,\rm p}$) are found by setting time and space
  derivatives to zero in (\ref{mean-field}). The mean-field
  steady-state equations then reduce to
\begin{equation}
  \forall i\qquad  \sum_{j\neq i}K_{ji}(\rho_j^{\rm P},\rho_i^{\rm P})=0.
  \label{equilplat}
\end{equation}
Eq.~\eqref{equilplat} simply states that, for each site of lane $i$,
the mean transverse flux of particles coming from all other lanes $j\neq i$ is
equal to the mean transverse flux leaving lane $i$. 
\begin{figure}[t]
  \centering
  \includegraphics{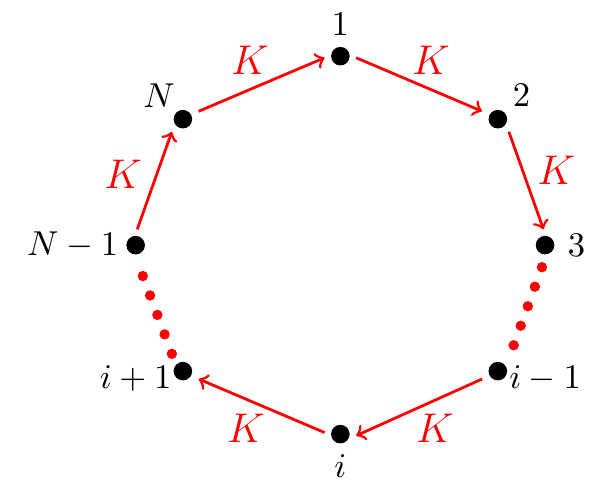}
  \caption{In a system with $N$ parallel lanes, organised on a ring,
    equilibrated plateaux can be realized with  a steady loop
    of non-zero transverse current $K>0$.}
  \label{fig:Kloop}
\end{figure}

If there are only two lanes, Eq.~\eqref{equilplat} implies
\begin{equation}
  K_{12}=K_{21}=0, 
  \label{equil2lanes}
\end{equation}
so that there is no net transverse current. However,
for a number of lanes $N\geq3$ the condition (\ref{equilplat})
can be satisfied by
the presence of a non-zero steady loop of transverse current. For
example, for $N=3$, with $K_{12}=K_{23}=K_{31}=K\neq0$, a
non-zero steady transverse current is present (see Fig~\ref{fig:Kloop} for an
illustration with $N$ lanes on a ring).

\subsection{Dynamical stability of equilibrated plateaux}
\label{sec:dynstab}
The plateaux solutions are only relevant to the steady state if they
are dynamically stable.  In~\ref{app:dynstab} we show, through a
dynamical linear stability analysis, that a small perturbation $\delta
\rho_i(x,t)$ around the equilibrated plateau solution:
\begin{equation}\label{FEpert}
  \rho_i(x,t)=\rho_i^{\,\rm p}+\sum_q\delta\rho_i^q(t)\exp(iqx)\;,
\end{equation}
where $q=2\pi n$ with $n=1,\ldots,\ell-1$,
vanishes exponentially rapidly in time if one considers systems
in which
\begin{equation}
  K_{ij}^i\equiv \partial_{\rho_i}K_{ij}(\rho_i^{\,\rm p},\rho_j^{\,\rm p})>0\quad\text{and}\quad K_{ij}^j\equiv \partial_{\rho_j}K_{ij}(\rho_i^{\,\rm p},\rho_j^{\,\rm p}) <0.
  \label{fluxesEP}
\end{equation}
Note that this is slightly weaker than condition~\eqref{fluxes}
since~\eqref{fluxesEP} only has to hold for equilibrated plateaux densities.

\subsection{Equilibrated and non-equilibrated reservoirs}
\label{sec:EqnonEqres}

\begin{figure}
  \begin{center}
    \includegraphics{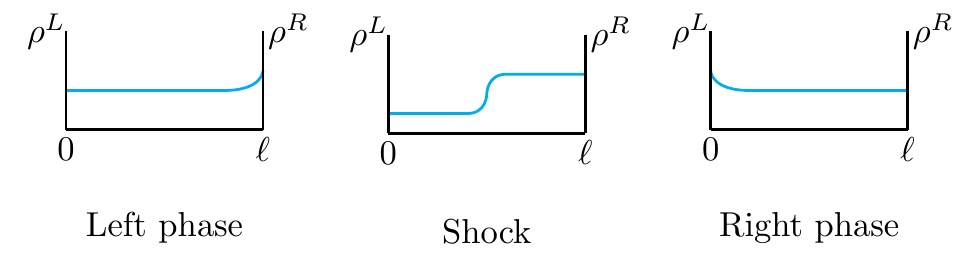}
  \end{center}
  \caption{Example of Left, Right and Shock phases}
  \label{fig:RLSphase}
\end{figure}

In analogy with equilibrated plateaux, one can define \emph{equilibrated reservoirs}, whose
densities satisfy
\begin{equation}
  \forall i \qquad \sum_{j\neq i}K_{ji}(\rho_j^{R,L},\rho_i^{R,L})=0,
  \label{equilBC}
\end{equation}
where $\rho_i^{L}$ and $\rho_i^{R}$ correspond to the densities of the
reservoirs at the left and right ends of lane $i$, respectively.  If
equilibrated reservoirs are imposed at both ends of the system, with
$\rho_i^R=\rho_i^L$, a simple solution of the mean-field equations in
steady state is found by connecting the reservoirs through constant
plateaux. Even though these boundary conditions are exceptional,
equilibrated plateau solutions play an important role in the steady
state of driven diffusive systems. As we show, far from the
boundaries, in a large system, the density profile is typically
constant.  With this in mind, in the following, we will use the term ``plateaux'' to
describe not only completely constant profiles, where the density is
the same on all sites of the system, but also flat \textit{portions}
of density profiles, see Fig.~\ref{fig:RLSphase}, which can be
connected to reservoirs or other density plateaux by small
non-constant boundary layers.

As noted previously, when equilibrated reservoirs are imposed at the
two ends of the system with the {\em same } densities
$\rho_i^L=\rho_i^R$, one observes dynamically stable plateaux with
$\rho_i(x)=\rho_i^L$.  Two other more general classes of boundary
condition are relevant.

\begin{figure}[ht]
  \begin{center}
    \includegraphics[scale=.92]{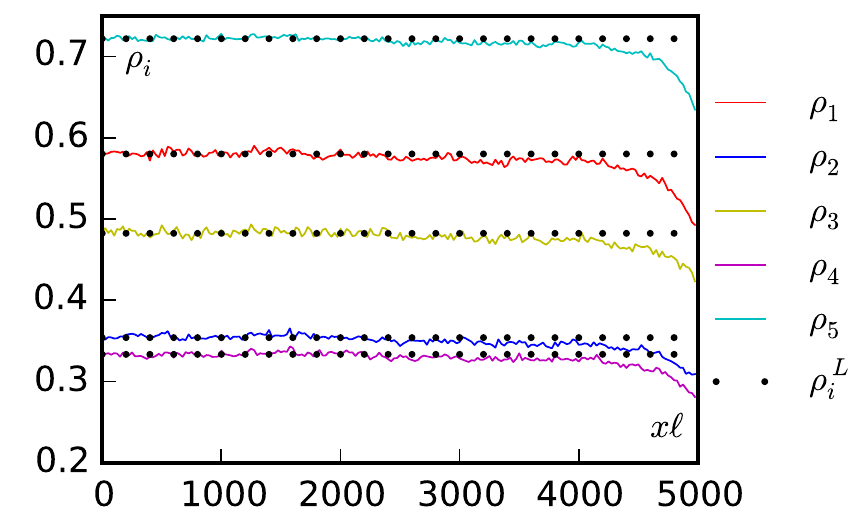}\hspace{1cm}\raisebox{2.2cm}{ \begin{tabular}{|
          c | c | c |}
                \hline
                $i$ & $\rho_i^L$& $\rho_i^R$\\
                \hline
                1&0.58&0.45\\
                2&0.35&0.25\\
                3&0.48&0.36\\
                4&0.33&0.24\\
                5&0.72&0.61\\
                \hline
              \end{tabular}}
\includegraphics[scale=.92]{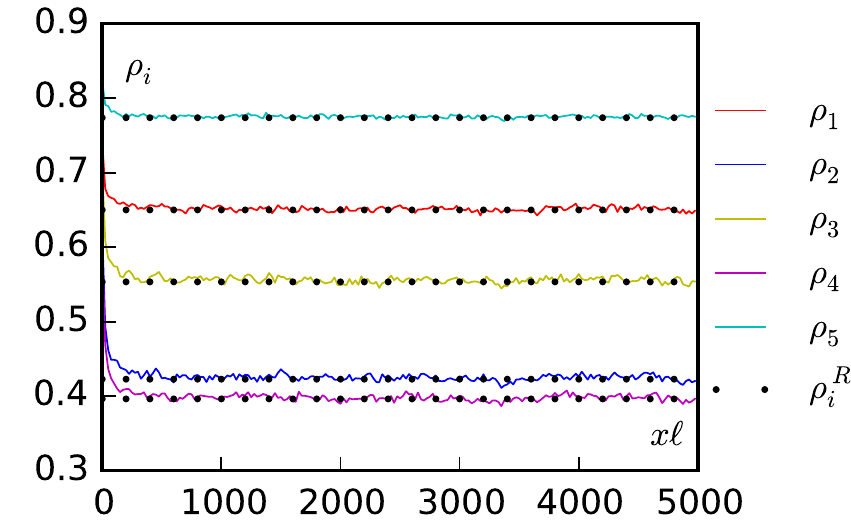}            \hspace{1cm}\raisebox{2.2cm}{ \begin{tabular}{|
          c | c | c |}
                \hline
                $i$&$\rho_i^L$& $\rho_i^R$\\
                \hline
                1&0.85&0.65\\
                2&0.69&0.42\\
                3&0.79&0.55\\
                4&0.65&0.40\\
                5&0.90&0.77\\
                \hline
              \end{tabular}}
            \caption{Connections of equilibrated plateaux to
              equilibrated reservoirs, for five parallel TASEPs as in
              Fig.~\ref{fig:Nlanes}. Top and bottom figures correspond
              to Left and Right phases. {\bf Top:} The density in the
              bulk is controlled by the left reservoirs; a small
              change of the right reservoir densities leaves the bulk
              plateau unchanged. {\bf Bottom:} The density in the bulk
              is controlled by the right reservoirs; a small change of
              the left reservoir densities leaves the bulk plateau
              unchanged.  For both cases, $p=1$, $d_{12}=10^{-3}$,
              $d_{23}=9.10^{-3}$, $d_{34}=2.10^{-3}$,
              $d_{45}=8.10^{-3}$, $d_{51}=3.10^{-3}$,
              $d_{15}=7.10^{-3}$, $d_{21}=4.10^{-3}$,
              $d_{32}=6.10^{-3}$, $d_{43}=5.10^{-3}$,
              $d_{54}=2.10^{-3}$ and results are obtained using
              continuous time Monte-Carlo simulations. }
    \label{fig:equilibratedreservoirs}
  \end{center}
\end{figure}

First, left and right equilibrated reservoirs can imply different sets
of plateaux densities in the bulk; how the density is then selected in
the bulk is one of the central questions tackled in this article. In
Fig.~\ref{fig:equilibratedreservoirs}, we show examples of Left and
Right phases, where the bulk density is controlled respectively by
left and right reservoirs. In a left (resp. right) phase, a small
modification of the right (resp. left) reservoir densities leaves the
bulk part of the density profile unaffected; only the boundary layer
connecting to the right (resp. left) reservoir changes.

\begin{figure}[ht]
  \begin{center}
    \includegraphics[scale=.92]{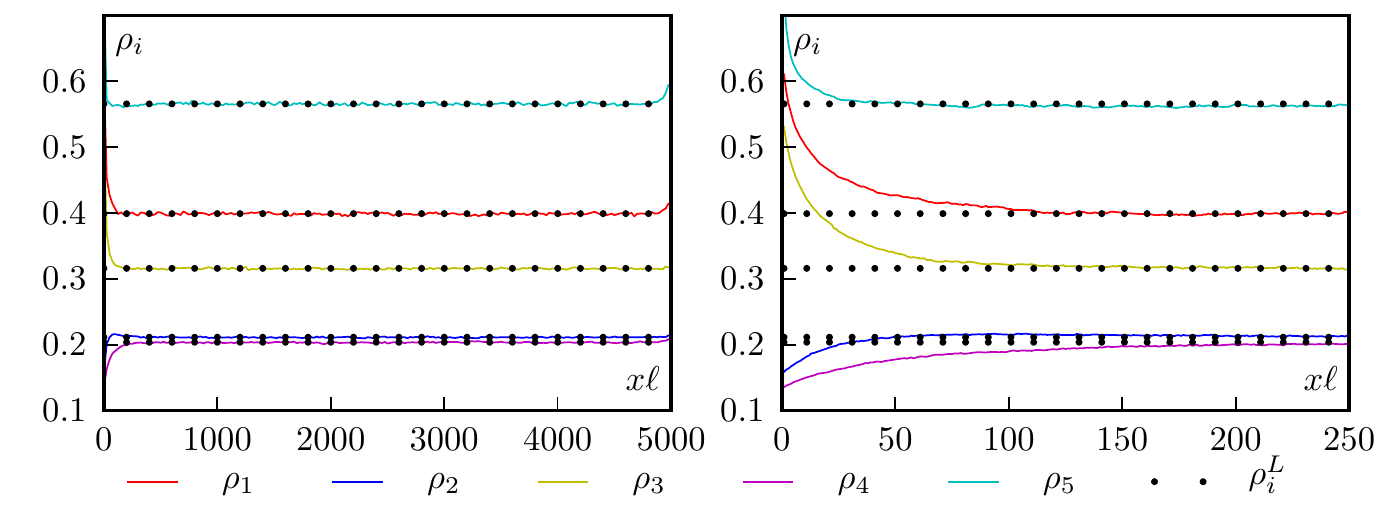}\raisebox{2.7cm}{ \begin{tabular}{|
          c | c | c |}
                \hline
                $i$ & $\rho_i^L$& $\rho_i^R$\\
                \hline
                1&0.68&0.45\\
                2&0.15&0.25\\
                3&0.58&0.36\\
                4&0.13&0.24\\
                5&0.92&0.61\\
                \hline
              \end{tabular}}
    \includegraphics[scale=.92]{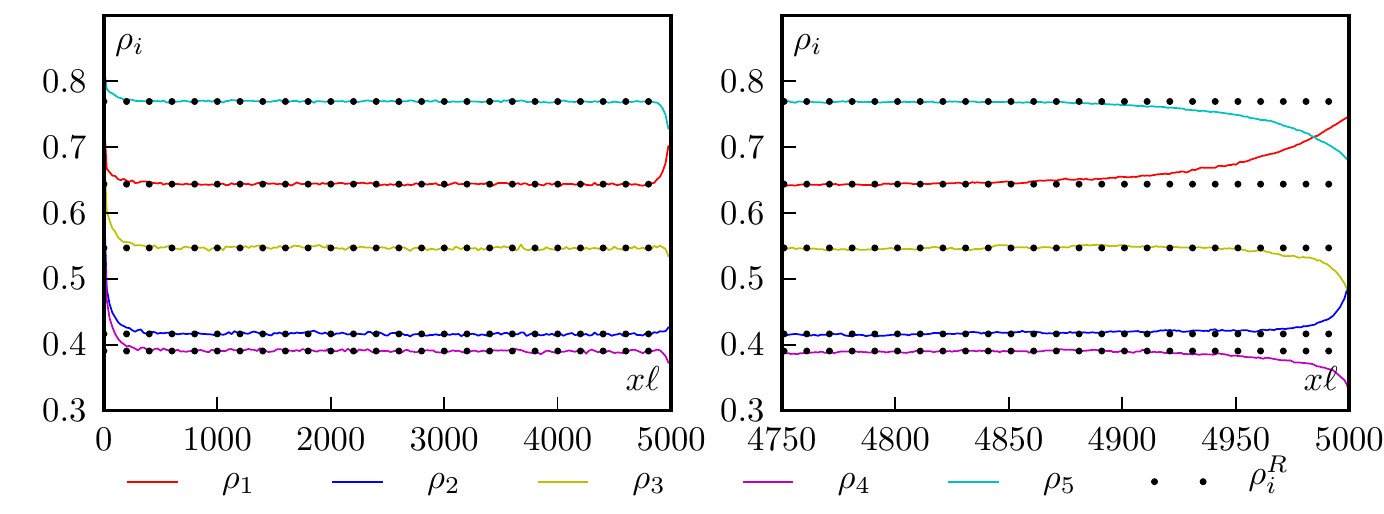}\raisebox{2.7cm}{ \begin{tabular}{|
          c | c | c |}
                \hline
                $i$&$\rho_i^L$& $\rho_i^R$\\
                \hline
                1&0.85&0.75\\
                2&0.69&0.52\\
                3&0.79&0.45\\
                4&0.65&0.30\\
                5&0.90&0.67\\
                \hline
              \end{tabular}}
            % QUESTION: Should this only arrives in the conclusion
            % when we mention unequilibrated plateaux? IS the text
            % enough to understand unequilibrated reservoirs?
            \caption{Connections of equilibrated plateaux to
              equilibrated and unequilibrated reservoirs, for five
              parallel TASEPs as in Fig.~\ref{fig:Nlanes}.  {\bf
                Top-Left:} A bulk plateau is connected to an
              equilibrated reservoirs on the right and an
              unequilibrated on the left. This is a left phase and the
              the bulk density is insensitive to small modification of
              the density of the right reservoirs. There are thus two
              boundary layers: one connecting the bulk plateau to the
              left unequilibrated reservoir, which controls the bulk
              density, and one connecting the bulk plateau to the
              right reservoir. {\bf Top-Right:} Close-up on the
              unequilibrated left reservoir. {\bf Bottom:} The same
              with left equilibrated reservoirs and right
              unequilibrated reservoirs, in a Right phase. For both
              cases, $p=1$, $d_{12}=10^{-3}$, $d_{23}=9.10^{-3}$,
              $d_{34}=2.10^{-3}$, $d_{45}=8.10^{-3}$,
              $d_{51}=3.10^{-3}$, $d_{15}=7.10^{-3}$,
              $d_{21}=4.10^{-3}$, $d_{32}=6.10^{-3}$,
              $d_{43}=5.10^{-3}$, $d_{54}=2.10^{-3}$ and results were
              obtained using continuous time Monte-Carlo simulations.}
%%             {\footnotesize      \begin{tabular}{| c | c c c c c |}
%%                 \hline
%%                 $i$ & 1 & 2 & 3 & 4 & 5 \\ 
%%                 \hline
%% %                $p_i$ & 1 & 1 & 1 & 1 & 1 \\
%% %                $d_{i,i+1}$ & 0.001 & 0.009 & 0.002 & 0.008 & 0.003 \\ 
%% %                $d_{i,i-1}$ & 0.007 & 0.004 & 0.006 & 0.005 & 0.002 \\
%%                 $\rho_i^L$ & 0.68 & 0.15 & 0.58 & 0.13 & 0.92 \\
%%                 $\rho_i^R$ & 0.45 & 0.25 & 0.36 & 0.24 & 0.61 \\
%%                 %% $\rho_i^B$ & 1 & 1 & 1 & 1 & 1 \\
%%                 \hline
%%               \end{tabular}
%%               \begin{tabular}{| c | c c c c c |}
%%                 \hline
%%                 $i$ & 1 & 2 & 3 & 4 & 5 \\ 
%%                 \hline
%% %                $p_i$ & 1 & 1 & 1 & 1 & 1 \\
%% %                $d_{i,i+1}$ & 0.001 & 0.009 & 0.002 & 0.008 & 0.003 \\ 
%% %                $d_{i,i-1}$ & 0.007 & 0.004 & 0.006 & 0.005 & 0.002 \\
%%                 $\rho_i^L$ & 0.85 & 0.69 & 0.79 & 0.65 & 0.90 \\
%%                 $\rho_i^R$ & 0.75 & 0.52 & 0.45 & 0.30 & 0.67 \\
%%                 %% $\rho_i^B$ & 1 & 1 & 1 & 1 & 1 \\
%%                 \hline
%%             \end{tabular}}
            \label{fig:unequilibratedreservoirs}
  \end{center}
\end{figure}

Second, the reservoirs can be unequilibrated, that is their densities
do not obey (\ref{equilBC}). In this case both left and right
reservoir densities are connected to bulk equilibrated plateaux by
small boundary layers. In Fig.~\eqref{fig:unequilibratedreservoirs},
we show examples of the corresponding left and right phases. Again, in
a left (resp. right) phase, a small modification of the right
(resp. left) reservoir densities leaves the bulk part of the density
profile unaffected; only the boundary layers connecting to the right
(resp. left) reservoir changes.

In this article, we focus on how the densities of the equilibrated
plateaux in the bulk are selected by the boundary conditions imposed
by equilibrated reservoirs and only comment on unequilibrated
reservoirs in the conclusion. The boundary layers connecting reservoir
to plateaux (or plateaux to other plateaux when a shock is observed as
in Fig~\ref{fig:RLSphase}) are beyond the scope of our work, but can
be studied using asymptotic
methods~\cite{Mukherji09,YSM12,Gupta2014PRE}.

\section{Phase diagram of multilane systems---a linear analysis}

The mean-field dynamics (\ref{mean-field}) evolves the density profiles towards flat
plateaux which are connected at their ends either to other plateaux or
to reservoirs. The acceptable plateaux solutions can thus be worked
out by considering which types of stationary profiles can connect a
bulk plateau to other densities at its right and left ends. For
instance, in the bottom panels of
Fig.~\ref{fig:unequilibratedreservoirs}, the steady-state solution of
the mean-field equations is made up of a plateau connected on its left
end to equilibrated reservoirs and on its right end to unequilibrated
reservoirs. In practice, we only study whether such profiles can exist
by looking at the vicinity of the equilibrated plateau. Namely, rather
than solving the full non-linear problem, we carry out a linear
analysis of the steady-state mean-field equations
\begin{equation}\label{eq:ssmf}
0=\partial_x[D_i\partial_x\rho_i(x,t)-J_i(\rho_i(x,t))]+\sum_{j\neq i}K_{ji}(\rho_j(x,t),\rho_i(x,t)),
\end{equation}
around equilibrated plateaux.

We thus look for steady-state profiles of the form
\begin{equation}
 \vert\rho(x))=\vert\rho^{\,\rm p})+\vert\delta\rho(x)),\qquad\text{where}\qquad \vert \rho(x)) \equiv \smallmatrix{\rho_1(x)\cr\vdots\cr\rho_N(x)}
\label{rhox}
\end{equation}
is a compact notation for $N$-dimensional vectors with components $\rho_i(x)$
 and $|\rho^{\, \rm
  p})$ corresponds to a set of equilibrated plateau densities.  Since
we have already established the \textit{dynamical} stability of
equilibrated plateaux in Section~\ref{sec:dynstab} (and~\ref{app:dynstab}), we are now only
concerned with the existence of \textit{spatial} stationary
perturbations $|\delta \rho(x))$ around them. In the following, we
simply refer to $|\delta \rho(x))$ as ``perturbations'', omitting
their implicit stationarity.

Linearising the steady-state mean-field equations~\eqref{eq:ssmf}
close to $|\rho^{\,\rm p})$ then yields
\begin{equation}
  0  =  D_i\partial_{xx}\delta\rho_i(x)-J_i^i(\rho_i^{\,\rm p})\partial_x\delta\rho_i(x)+\sum_{j\neq i}[K_{ji}^i\delta\rho_i(x)+K_{ji}^j\delta\rho_j(x)],
  \label{SSMF}
\end{equation}
where   $J_i^i = \frac{\partial J_i(\rho_i)}{ \partial \rho_i}$
and $K_{ji}^i$, $K_{ij}^j$ are defined in (\ref{fluxesEP}).
Equation (\ref{SSMF})
can be seen as a first  order ordinary differential equation
$\partial_x \vert v\rangle=M\vert v\rangle$, where
\begin{equation}
  M(\rho_1^{\,\rm p},\dots,\rho_N^{\,\rm p}) =  \left(\begin{array}{ccc|ccc}
      \frac{J^1_1}{D_1} & & & & & \\
      & \ddots & & & -M^K & \\
      & & \frac{J^N_N}{D_N} & & & \\ \hline
      \frac 1 D_1 & & & & & \\
      & \ddots & & & 0 & \\
      & & \frac 1 D_N & & &
    \end{array}\right),\; 
  \vert v\rangle=\smallmatrix{D_1\delta\rho_1'(x) \cr \vdots \cr D_N\delta\rho_N'(x) \cr \delta\rho_1(x) \cr \vdots \cr \delta\rho_N(x)}
  \label{stabmatN}
\end{equation} 
Here, $M^K$ is an $N\times N$ matrix defined as
\begin{equation}
  M^K_{ii}=\sum_{j\neq i}K_{ji}^i \qquad M^K_{ij}=K_{ji}^j\,,
  \label{MK}
\end{equation}
$|v\rangle$ is a compact notation for $2N$-dimensional
vectors and $\delta\rho_i'(x)\equiv \partial_x \delta\rho_i(x)$. When $M$ is diagonalizable, solutions of this ordinary
differential equation are of the form
\begin{equation}
  \label{eq:diag}
  \vert v(x)\rangle=\sum_{k=0}^{2N-1}\alpha_k e^{\lambda_k x} \vert v^k\rangle
\end{equation}
where $\vert v^k\rangle=(D_1 \lambda_k
\delta\rho^k_1,\dots,D_N \lambda_k\delta\rho^k_N,\delta\rho_1^k,\dots,\delta\rho_N^k)$
are the $2N$ eigenvectors of the matrix $M$ and $\lambda_k$ the
corresponding eigenvalues. This implies that the perturbation to the density profile  (\ref{rhox})  may be decomposed as
\begin{equation}
  \label{eqn:EVdecomp}
\vert \delta \rho(x) ) = \sum_{k=0}^{2N-1} \alpha_k e^{\lambda_k x} \vert \delta\rho^k)\;.
\end{equation}

The study of the spectrum of the matrix $M(\rho_1^{\,\rm
  p},\dots,\rho_N^{\,\rm p})$ reveals whether given boundary
conditions can be connected to a set of bulk plateaux of densities
$|\rho^{\,\rm p})$ through appropriate perturbations $\vert \delta
\rho(x) )$ (\ref{eqn:EVdecomp}). This will allow us in section \ref{LSA:phasediag} to construct
the phase diagram.

\begin{figure}[H]
  \centering
  \includegraphics{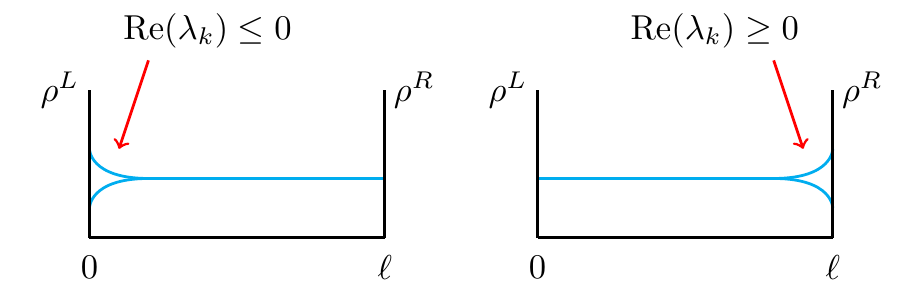}
  \caption{Connecting a reservoir to an equilibrated  plateau.
     With $\Re(\lambda_k)\le0$, one can construct perturbations
    connecting an equilibrated plateau to the left reservoirs. With
    $\Re(\lambda_k)\ge0$, one can construct perturbations connecting an
    equilibrated plateau to right reservoirs.}
  \label{fig:lambdas2}
\end{figure}

We now detail how the spectrum of $M$ is organised and the role played
by its eigenvectors. One might expect the study of the spectrum of a
$2N$ dimensional matrix to be rather complex. But, as we show below,
a simple result for the phase diagram, which is solely
determined by the \textit{number} of eigenvalues with positive (or
negative) real part, can be obtained. This can be found using a simple
criterion which does not require knowledge of the full spectrum. 

Specifically, we show in section~\ref{sec:lambda0} that there is
always one zero eigenvalue, $\lambda_0$, whose eigenvector corresponds
to a (unique) uniform perturbation which preserves the equilibrated
plateau condition. Then, in section~\ref{sec:EVEqPlat}, we show that
the rest of the spectrum can be organised in two different ways:
Either with $N-1$ eigenvectors with $\Re(\lambda_k)<0$ and $N$ with
$\Re(\lambda_k)>0$, or with $N$ eigenvectors with $\Re(\lambda_k)<0$
and $N-1$ with $\Re(\lambda_k)>0$. We show that the first case is
compatible with a Left Phase and the second with a Right Phase (see
Fig. \ref{fig:lambdas2}) .  The separation between these two cases is
thus controlled by the change of sign of the real part of an
eigenvalue, which we show in section~\ref{sec:EMPEV} to occur at
extrema of the total current $J_{\rm tot}$ within the space of
equilibrated plateaux.

Sections~\ref{sec:lambda0} to~\ref{sec:EMPEV} thus establish the
possibility, at the linear level, of a non-uniform profile $\vert
\rho(x) )$ connecting a plateau $\vert \rho^{\, p})$ to other
equilibrated density vectors at one of its ends. By studying the
dynamics of the shocks connecting the plateaux, in
section~\ref{LSA:phasediag}, we derive a simple recipe for
constructing the phase diagram.  Readers who are primarily interested
in the construction of the phase diagram may thus proceed directly to
Sec.~\ref{LSA:phasediag} upon a first reading of this paper.

\subsection{Uniform shift of equilibrated plateaux}
\label{sec:lambda0}
We now show that
the matrix $M$ given in (\ref{stabmatN}) always admits a unique $\lambda_0=0$ eigenvalue,
associated to uniform shifts of the densities of equilibrated plateaux.
Using $M|v^0\rangle=0$ implies that
$|v^0\rangle=(0,\dots,0,\delta\rho_1^0,\dots,\delta\rho_N^0)$ with
\begin{equation}
  M^K\vert\delta\rho^0) = 0.
  \label{unique}
\end{equation} 
One easily checks that $M^K$ is a Markov (or Intensity) matrix by
summing its row elements: $\forall j,\; \sum_i M^K_{ij}=0$. By the
Perron-Frobenius theorem $M$ has a unique $\lambda=0$ eigenvector
$|\delta \rho^0)$ which is real and has all its non-zero components of
the same sign (which we take to be positive).

It is easy to check that a
perturbation of the density vector from $|\rho^{\, \rm p})$ to  $|\rho^{\, \rm p}) + \eps |\delta \rho^0)$  for small $\epsilon$  indeed results in an  
equilibrated plateau, since to first order in $\epsilon$
\begin{eqnarray}  \sum_{j\neq i} K_{ji}(\rho_j^{\,\rm p}+\eps \delta \rho_j^0,\rho_i^{\,\rm p}+\eps\delta \rho_i^0)&=  \sum_{j\neq i} K_{ji}(\rho_j^{\,\rm p},\rho_i^{\,\rm p})+\eps\sum_{j\neq i} K_{ji}^i\delta \rho_i^0+\eps\sum_{j\neq i} K_{ji}^j\delta \rho_j^0\nonumber\\
  &=  0+\eps \sum_j M_{ij}^K  \delta\rho^0_j  =0\;,
\end{eqnarray}
where we used the fact that $\sum_{j\neq i}
K_{ji}(\rho_j^{\,\rm p},\rho_i^{\,\rm p})=0$ since $|\rho^{\,\rm p})$ is
equilibrated. 

The vector $|\delta \rho^0)$ is therefore the unique tangent vector to the
one-dimensional manifold of equilibrated plateaux: all infinitesimal
perturbations $|\delta \rho^{\,\rm p})$ such that $|\rho^{\,\rm
  p})+|\delta \rho^{\,\rm p})$ remains equilibrated are thus parallel
to $|\delta \rho^{\,\rm 0})$. It is important to bear in mind that
since $M$ is a function of $|\rho^{\rm p})$, $|\delta \rho^0)$ also
depends on $|\rho^{\, \rm p})$.
% so that the manifold of equilibrated plateaux is not a straight
% line.

\subsection{Connecting equilibrated plateaux to reservoirs}

\label{sec:EVEqPlat}
In this subsection we consider how a bulk plateau density vector
$|\rho^{\, \rm p})$ may be connected by a density profile to
equilibrated reservoirs $|\rho^{\,\rm r})$. For the sake of clarity, we
focus here on the main results whose detailed derivations are given
in~\ref{sec:appspectrum}.

First, as noted above, we show in~\ref{sec:appspectrum1} that in
addition to the $\lambda=0$ eigenvalue discussed in the previous
subsection, the spectrum of $M$ is composed of either
\begin{description}
\item[$(i)$]
 $N$ eigenvalues with
positive real parts and $N-1$ eigenvalues with negative real parts, 
\item[$(ii)$]
 $N{-}1$ eigenvalues with positive real parts and $N$ eigenvalues
with negative real parts.
\end{description}
We then show in~\ref{sec:appspectrum2} that in case $(i)$ the plateau
density vector can only be connected to arbitrary equilibrated
reservoir densities to the right.  Thus, for equilibrated reservoirs,
such plateaux can only be observed in a Left Phase---a phase in which
the bulk plateaux are controlled by the left reservoirs---with
$|\rho^{\, \rm p})=|\rho^L)$. Conversely, in case $(ii)$ the plateau
density vector can only be connected to arbitrary equilibrated
reservoir densities to the left. Such plateaux can thus only be
observed in a Right Phase.

As the density vector $|\rho^{\, \rm p})$ varies, the transition between the
two cases  corresponds to the vanishing of an eigenvalue, whose
real part will  then change sign. Precisely at the transition one then has $N-1$ eigenvalues with $\Re(\lambda_k)>0$,  $N-1$ eigenvalues with
$\Re(\lambda_k)<0$ and two zero eigenvalues. Since the
eigenspace associated with $\lambda=0$ is of dimension 2 but there is
only one $\lambda=0$ eigenvector, the matrix $M$ is then non-diagonalizable. Let us now show that this
corresponds to the local extrema of $J_{\rm tot}$ within the space of
equilibrated plateaux.

\begin{figure}
  \centering
  \includegraphics{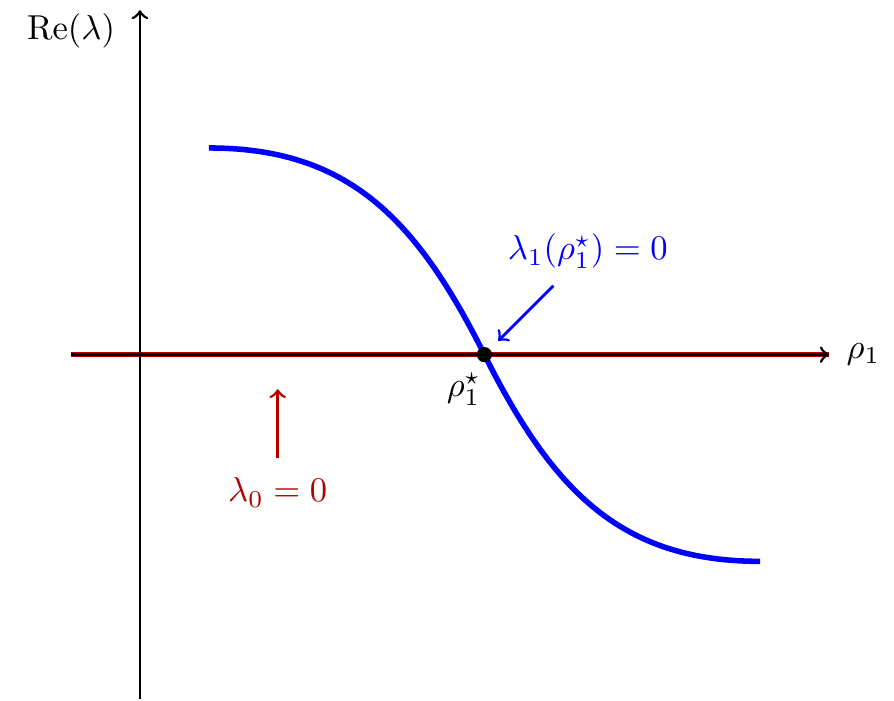}
  \caption{
Change of sign of an eigenvalue of stability matrix  $M$ given by (\ref{stabmatN}). The eigenvalue $\lambda_1$ vanishes at $\rho_1=\rho_1^\star$ while $\lambda_0=0$ everywhere.}
  \label{fig:change-sign}
\end{figure}

\subsection{Change of sign of $\Re(\lambda)$ and extrema of the current}
\label{sec:EMPEV}
From the discussion above it is clear that for the phase diagram to be
richer than simply one Left Phase or one Right Phase, we need an
eigenvalue, let us denote it $\lambda_1$, to vanish and change sign as
the reservoir densities are varied.

For equilibrated plateaux $|\rho)$ (to lighten the notation we drop
the $\rm p$ superscript), all densities $\rho_i$ and all
eigenvalues of $M(\rho_1,\dots,\rho_N)$ can be written as functions of
the density of the first lane $\rho_1$ via the equilibrated plateau
condition~\eref{equilplat}. For some value $\rho_1^\star$, we thus
require $\lambda_1(\rho_1^\star)=0$ with
$\lambda_1(\rho_1^\star\pm\varepsilon)\neq0$ (see
Fig.~\ref{fig:change-sign}). As we show below, this happens at all the local extrema of the total
current
\begin{equation}
  J_{\rm tot}(\rho_1)\equiv\sum_i J_i[\rho_i(\rho_1)]
\label{Jtot}
\end{equation}
within the set of equilibrated plateaux.

At a given plateau $|\rho)$ , the tangent vector to the set of
equilibrated plateaux is the $\lambda=0$ eigenvector $|\delta\rho^0)$
of matrix $M^K(\rho_1,\dots,\rho_N)$, defined in (\ref{MK}).  The
extrema of $J_{\rm tot}$ within the set of equilibrated plateaux thus
occur when
\begin{equation}
  \label{ecp}
  \mbox{d}J_{\rm tot}= \nabla J_{\rm tot} \cdot |\delta \rho^0)=\sum_iJ_i^i\delta\rho^0_i=0 \;.
\end{equation}

The linearized
steady-state mean-field equation~(\ref{SSMF}) becomes, when we take
$\delta \rho_i = \delta \rho_i^1 {\rm e}^{\lambda_1 x}$, an
eigenvector equation for $|\delta \rho^1)$ which reads
\begin{equation}
  (\lambda_1)^2D_i\delta\rho^1_i-\lambda_1J^i_i\delta\rho^1_i+ \sum_{j\neq i}(K_{ji}^i\delta\rho^1_i+K_{ji}^j\delta\rho^1_j)=0.
  \label{0sbatta}
\end{equation}
By summing~(\ref{0sbatta}) over $i$, one finds  $(\lambda_1)^2\sum_i
D_i\delta\rho^1_i-\lambda_1\sum_i J^i_i\delta\rho^1_i=0$ and thus
\begin{equation}
   \sum_iJ_i^i\delta\rho^1_i= \lambda_1\sum_i D_i \delta\rho_i^1.
  \label{current12}
\end{equation}
As we approach $\rho_1^\star$, where  $\lambda_1 \to 0$,   the left hand side
of (\ref{current12}) must tend to zero.

However, since $M^K$ is a Markov matrix there is a unique eigenvector
corresponding to eigenvalue zero which is $|\delta\rho^0)$.  Therefore
as $\lambda_1\to 0$, we must have $\delta\rho^1_i \to \delta\rho^0_i$.
Thus as $\rho_1 \to \rho^\star_1$ we conclude that
\begin{equation}
\sum_iJ_i^i\delta\rho^1_i \to \sum_iJ_i^i\delta\rho^0_i = {\rm d} J_{\rm tot} \to 0.
\label{current}
\end{equation}

As $\rho_1 \to \rho_1^\star$ this corresponds to the extremal current condition, Eq. (\ref{ecp}). The vanishing of $\Re(\lambda_1)$
thus occurs at the extrema of $J_{\rm tot}$.

Finally, since $|\delta \rho^0)$ is the unique zero eigenvector of a
Markov matrix $M^K$, all its components are of the same sign.  As
$D_i$ are all positive, equations~\eref{current12} and \eref{current}
imply that $\Re(\lambda_1(\rho_1))$ and $J'_{\rm tot}(\rho_1)$ are of
the same sign in the vicinity of the extrema of $J_{\rm tot}$. Since
neither of them vanish and change sign elsewhere,
$\Re(\lambda(\rho_1))$ and $J'_{\rm tot}(\rho_1)$ must always have the
same sign.

\subsection{Phase diagram and extrema of the currents}
\label{LSA:phasediag}
From sections~\ref{sec:EVEqPlat} (\ref{sec:appspectrum}) and
\ref{sec:EMPEV}, one concludes that for equilibrated plateaux density
vectors $|\rho)$, the sign of $J'_{\rm tot}(\rho_1)$ controls the
boundary reservoirs to which the plateaux can be connected:
\begin{itemize}
  \itemsep-.2cm
\item For $J'_{\rm tot}(\rho_1) >0$, the plateaux can only be
  connected to different equilibrated densities at their right ends\\
\item For $J'_{\rm tot}(\rho_1) <0$, the plateaux can only be
  connected to different equilibrated densities at their left ends.\\
\item $J'_{\rm tot}(\rho_1)=0$ is a singular limiting case
  between these two regimes;
  \begin{itemize}
  \item[-] For a local maximum of $J_{\rm tot}$, perturbations with
    $\delta\rho_1>0$ are in a region where $J'_{\rm tot}(\rho_1) <0$ and
    can thus connect to equilibrated densities with $\tilde
    \rho_1>\rho_1$ on the left. Perturbations with $\delta\rho_1<0$ are
    in a region where $J'_{\rm tot}(\rho_1) >0$ and can thus connect to
    equilibrated densities with $\tilde \rho_1<\rho_1$ on the right.
    
  \item[-] For a local minimum of $J_{\rm tot}$, perturbations with
    $\delta\rho_1>0$ are in a region where $J'_{\rm tot}(\rho_1) >0$ and
    can thus connect to equilibrated densities with $\tilde
    \rho_1>\rho_1$ on the right. Perturbations with $\delta\rho_1<0$ are
    in a region where $J'_{\rm tot}(\rho_1) <0$ and can thus connect to
    equilibrated densities with $\tilde \rho_1<\rho_1$ on the left.
  \end{itemize}
  The exact shape of the profile in this case, {which may involve algebraic functions rather than exponential}, is beyond the scope of
  this paper.
\end{itemize}

\begin{figure}
  \includegraphics{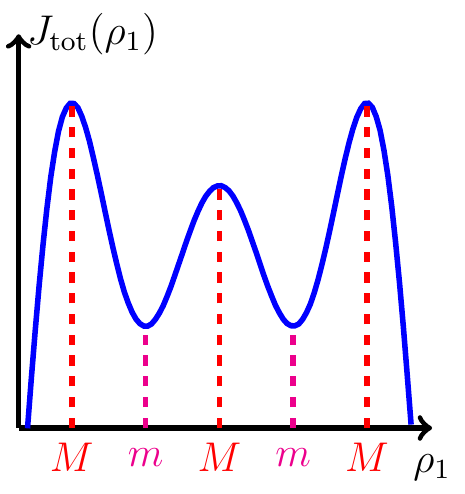}%\hspace{-.5cm}
  \includegraphics{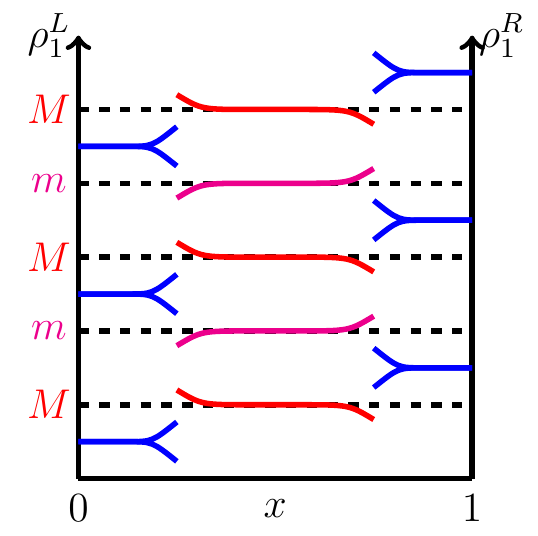}%\hspace{-.5cm}
  \includegraphics{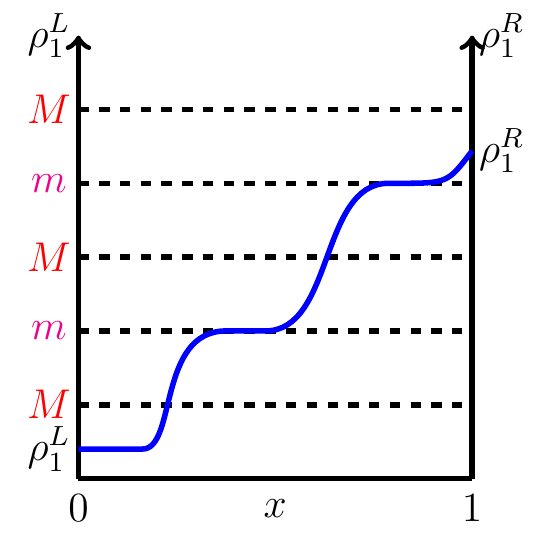}
  \caption{{\bf Left}: Example of current density relation $J_{\rm
      tot}(\rho_1)$. {\bf Center}: Linear stability of all
    equilibrated plateau. {\bf Right}: Example of possible profiles
    connecting left and right reservoirs. The shocks connecting the
    plateaux need not be stationary.}
\label{fig:extrema}
\end{figure}

To construct the phase diagram, one thus computes $J_{\rm tot}(\rho_1)$
and find the values of all its extrema, as shown in the left panel of
Fig.~\ref{fig:extrema}. This allows one to determine for each plateau
if, and how, it can be connected to other equilibrated densities at
its left or right ends, as shown in the central panel of
Fig.~\ref{fig:extrema}. Then, for given sets of equilibrated
reservoirs $|\rho^L)$ and $|\rho^R)$, one can construct the profiles
which are in agreement with the linear  analysis. As shown in
the right panel of Fig.~\ref{fig:extrema}, these profiles are
monotonous and made of successions of plateaux, whose densities are
set by the reservoirs or correspond to local extrema of the current,
connected by shocks.

\begin{figure}
  \begin{center}
    \includegraphics{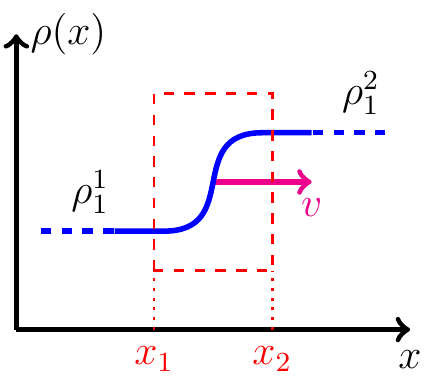}
  \end{center}
  \caption{Schematic of a shock profile connecting a plateau of
    density $\rho_1^1$ to a plateau of density $\rho_1^2$ moving with
    speed $v$.  The number of particles within the red rectangle varies at a rate
    $v(\rho^1_{\rm tot}-\rho^2_{\rm tot})$. If $v>0$ the plateau at
    $\rho_1=\rho_1^1$ spreads while the plateau at $\rho_1=\rho_1^2$
    recedes. Conversely, for $v<0$, the plateau at $\rho_1=\rho_1^2$
    spreads while the one at $\rho_1=\rho_1^1$ recedes.}
\label{fig:shockdynamics}
\end{figure}

While such shocks are in agreement with the linear analysis of the
steady-state mean-field equations, they need not be stationary. Let us
consider the situation depicted in Fig.~\ref{fig:shockdynamics}. The
rate of change of the number of particles in the region $[x_1,x_2]$
due to the moving shock profile is
\begin{equation}
  \dot {\cal N}_{[x_1,x_2]}= v\sum_i(\rho_i^1-\rho^2_i) \;,
\end{equation}
where $v$ is the shock velocity and the sum is over lanes. By the
continuity equation for the number of particles the rate of change is
also given by the fluxes at the boundary of the region, thus
\begin{equation}
  \dot {\cal N}_{[x_1,x_2]}=\int_{x_1}^{x_2} \sum_i \dot \rho_i = \sum_i \left[J_i(\rho^1_i)-J_i(\rho_i^2)\right]
\end{equation}
Equating the two expressions, the velocity $v$ of the shock is thus given by
\begin{equation}
  v=\frac{J_{\rm tot}(\rho^1_1)-J_{\rm tot}(\rho^2_1)}{\rho^1_{\rm tot}-\rho^2_{\rm tot}}
\end{equation}
where $\rho_{\rm tot}=\sum_i \rho_i$.

Let us now show that the sign of $\rho^1_{\rm tot}-\rho^2_{\rm tot}$
is the same as the sign of $\rho^1_{1}-\rho^2_{1}$. Since the tangent
vector to the set of equilibrated plateau, $|\delta \rho^0)$, has all
its components of the same sign, a small increase in $\rho_1$ results
in a new equilibrated plateau with small increases of $\rho_{i\neq
  1}$. The total density $\rho_{\rm tot}$ is thus an increasing
function of $\rho_1$: if one has $\rho_1^1<\rho_1^2$
(resp. $\rho_1^1>\rho_1^2$) then one must have $\rho_{\rm
  tot}^1<\rho_{\rm tot}^2$ (resp. $\rho_{\rm tot}^1>\rho_{\rm
  tot}^2$).  Increasing and decreasing shocks thus correspond to
$\rho^2_1>\rho^1_1$ and $\rho^2_1<\rho^1_1$.

Increasing shocks thus move to the right if $J_{\rm
  tot}(\rho_1)<J_{\rm tot}(\rho_2)$ and the plateau of density
$\rho_1$ invades the plateau of density $\rho_2$. Conversely if
$J_{\rm tot}(\rho_1)>J_{\rm tot}(\rho_2)$, the shock moves to the left
and the plateau of density $\rho_2$ widens. For decreasing shocks, on
the the other hand, the plateau corresponding to the larger current widens.

Note that the shocks on all lanes are co-localized, as shown by our
linear stability analysis. Intuitively, unequilibrated densities would
otherwise have to coexist between, say, the shock on lane $1$ and on
lane $2$, which would make the profiles unstable 
\cite{SARS10}.

\subsubsection{Construction of the phase diagram}
\label{sec:PDcons}
We can now gather everything together  in a simple rule for constructing the phase diagram:
\begin{itemize}
  \itemsep-0.3cm
\item For increasing profiles, when $\rho_1^L<\rho_1^R$, among all the
  plateaux in agreement with the linear analysis, the one
  corresponding to the smallest current spreads while the others
  recedes. The possible plateaux correspond either to the reservoir
  densities, if $J'_{\rm tot}(\rho_1)^L>0$ or $J'_{\rm
    tot}(\rho_1)^R<0$, or to local minima of the currents (local
  maxima correspond to decreasing profiles as shown in
  Fig.~\ref{fig:extrema}). For continuous $J_{\rm tot}(\rho_1)$, the
  global extrema of $J_{\rm tot}$ on $[\rho^L,\rho^R]$ are either at
  the boundaries or at local extrema so that the minima of $J_{\rm
    tot}$ among the possible plateaux correspond to the global minima
  of $J_{\rm tot}$ on $[\rho_1^L,\rho_1^R]$. Note that if there are
  several minima with the same value of the current, a shock
  connecting these densities does not propagate
  ballistically but simply diffuses.   \\
\item For decreasing profiles, when $\rho_1^L>\rho_1^R$, the converse
  reasoning leads to the selection of the plateau(x) with the largest
  current.\\ This is summarized in the generalized ECP discussed
  in section~\ref{sec:ECP}.
\end{itemize}

\subsubsection{A generalized extremal current principle}
\label{sec:ECP}
Let us now show how our approach relates to the extremal current
principle. As noted in Sec.~\ref{sec:recapTASEP}, based on the
assumption that there exists a relation $J(\rho)$ between the particle
current $J$ and local density $\rho$, an extremal current principle
can be used to derive the phase diagram of the TASEP~\cite{Krug}.
In~\cite{EKST11}, this principle was generalized to two-lane systems
with equilibrated reservoirs by extremising the {\em total} current
summed over the two lanes. The results of the previous section thus
generalizes this approach to the $N$-lane case in the presence of
transverse currents.

Given two sets of equilibrated reservoirs at the left and right ends
of the system, section~\ref{sec:PDcons} shows the equilibrated
densities $\rho_i^B$ observed in the bulk to satisfy
\begin{equation}
  J_{\rm tot}(\rho_1^B)=\left\{
    \begin{array}{lr}
      \max_{\rho_1^B\in[\rho_1^R,\rho_1^L]} \ J_{\rm tot}(\rho^B_1) \qquad \text{for} \quad \rho_1^L>\rho_1^R \\
      \min_{\rho^B_1\in[\rho_1^L,\rho_1^R]} \ J_{\rm tot}(\rho^B_1) \qquad \text{for} \quad \rho_1^L<\rho_1^R  \;. \\
    \end{array}
  \right.
  \label{extrcurrNL}
\end{equation}
Note that the $N=2$ case is rather atypical in that $K=0$
automatically, the only non-zero stationary currents are thus
longitudinal. For $N>2$, non-zero transverse currents are generically
present and it is perhaps surprising that they do not affect the
extremal current principle.

\section{Microscopic models of multilane systems}

\label{sec:numerics}

In the previous section we presented a linear stability analysis that
allowed us to predict the phase diagrams of multilane systems starting
from their hydrodynamic descriptions. We now compare our theoretical
predictions to various  microscopic models. In particular, we discuss the
validity of the generalised extremal current
principle~\eqref{extrcurrNL}.

We consider $N$ parallel lanes organised transversely ``on a ring'',
as in Fig.~\ref{fig:Kloop}: particles can hop from lane $i$ to lane
$i\pm1$, with periodic boundary conditions. This will allow us to
study the phenomenology of non-zero transverse currents. We first
consider $N=10$ parallel TASEPs in section~\ref{sec:simuTASEP} before
turning to a partial exclusion processes in
section~\ref{sec:simuTAPEP}. Finally, in section~\ref{sec:beyondMF} we
discuss a case where, while the simple mean-field approximation we employ is not exact,
the generalized extremal current principle still holds.

\subsection{$N$ TASEPs on a ring}
\label{sec:simuTASEP}

We consider a system of length $\ell$ composed of $N$ one-dimensional
TASEPs. Particles hop along each lane at constant rates and there can
be only one particle per site. Each lane is connected to reservoirs at
its ends and particles can hop from lane $i$ to the neighbouring lanes
$i\pm 1$, giving rise to transverse currents between the lanes. For
example, this could mimic the structure of microtubules which are
composed of several protofilaments along and between which molecular
motors can walk. Previously the case $N=2$ has been considered (see
for instance~\cite{EKST11} and references therein) whereas the $N=3$
case (without periodic boundary conditions in the transverse
direction) has been studied numerically~\cite{WJWW14}.  Here we show
that indeed the system of $N$ TASEPs for arbitrary $N$ may be
determined by the generalized ECP based on the total current.

A particle on lane $i$ at site $j$ can hop to  the neighbouring site
$j+1$ of lane $i$ with rate $p_i$ and on the site $j$ of lanes $i\pm 1$
with rate $d_{i,i\pm 1}$, provided the arrival site is empty. The
microscopic mean-field equations, within the approximation of factorising all 
density correlation functions, describing these $N$ parallel TASEPs
read
\begin{eqnarray}
  \frac{\mathrm{d}\rho_{i,j}}{\mathrm{d}t} &=\nonumber 
  p_i\,\rho_{i,j-1}(1-\rho_{i,j})-p_i\,\rho_{i,j}(1-\rho_{i,j+1})\\
%  &\quad+ q_i\,\rho_{i,j+1}(1-\rho_{i,j})-q_i\,\rho_{i,j}(1-\rho_{i,j-1})\\
  &\quad+d_{i-1,i}\,\rho_{i-1,j}(1-\rho_{i,j})-d_{i,i-1}\,\rho_{i,j}(1-\rho_{i-1,j})\nonumber \\
  &\quad+d_{i+1,i}\,\rho_{i+1,j}(1-\rho_{i,j})-d_{i,i+1}\,\rho_{i,j}(1-\rho_{i+1,j})
\end{eqnarray}
where $\rho_{i,j}=\langle n_{i,j}\rangle$ is the mean occupancy of the
site $j$ of lane $i$ and we use transverse periodic boundary
conditions by identifying lanes $i=N+1$ and $i=1$.

We can now define the mean-field currents of the microscopic models:
\begin{eqnarray}
  {\cal J}_{ij}=p_i\,\rho_{i,j}(1-\rho_{i,j+1})\\%-q_i \rho_{i,j+1}(1-\rho_{i,j}) \\ 
{\cal  K}_{(i,i+1)j}=d_{i,i+1}\,\rho_{i,j}(1-\rho_{i+1,j})-d_{i+1,i}\rho_{i+1,j}(1-\rho_{ij}).
\end{eqnarray}
where ${\cal J}_{ij}$ is the mean-field approximation of the
`longitudinal' particle current between sites $j$ and $j+1$ of lane
$i$ and ${\cal K}_{(i,i+1)j}$ is the mean-field approximation of the
`transverse' particle current between sites $j$ of lanes $i$ and
$i+1$.

To obtain coarse-grained equations, we introduce the rescaled position
$x=j/\ell$ along the lattices which now goes from $x=0$ to $x=1$.
One then uses a standard Taylor expansion assuming that $\rho_{i,j}$
varies slowly from site to site along a given lane:
\begin{equation}
  \rho_{i,j\pm1} \equiv \rho_i(x\pm \frac{1}{\ell}) \simeq \rho_i(x)\pm
  \frac 1 \ell \frac{\partial \rho_i(x)}{\partial x}+\frac {1}{2\ell^2}
\frac{\partial^2 \rho_i(x)}{\partial x^2}.
  \label{cont_limit}
\end{equation}
By keeping terms up to second order in gradients one can then write
the coarse-grained mean-field equations:
\begin{eqnarray}
  \dot{\rho}_i(x)=&-\partial_x[J_i(\rho_i(x))-D_i\partial_x\rho_i(x)]+
  \nonumber \\ &+K_{i-1,i}[\rho_{i-1}(x),\rho_i(x)]-
  K_{i,i+1}[\rho_i(x),\rho_{i+1}(x)]
  \label{eq:mean-field}
\end{eqnarray}
where 
\begin{equation}
  J_i(\rho_i(x))=\frac{p_i}{\ell}\rho_i(x)[1-\rho_i(x)],\quad\text{and}\quad -D_i\partial_x\rho_i=-\frac{p_i}{2\ell^2}\partial_x\rho_i
\label{eq:transcurrNTASEPS}
\end{equation}
are the advective and the diffusive parts of the mean-field current
along lane $i$. We stress again that in the  models we consider the longitudinal
current of one lane depends solely on the occupancies in that
lane. $K_{i,i+1}$ is the transverse current from lane $i$ to lane
$i+1$ defined as
\begin{equation} K_{i,i+1}[\rho_i(x),\rho_{i+1}(x)]=d_{i,i+1}\rho_i(x)[1-\rho_{i+1}(x)]-d_{i+1,i}\rho_{i+1}(x)[1-\rho_i(x)],
  \label{trcurr}
\end{equation}
with $K_{i,i+1}=-K_{i+1,i}$. In steady state, the equilibrated
plateau~\eqref{equilplat} condition then reads
\begin{eqnarray}
\forall i,\qquad K_{i,i+1}=-K_{i+1,i}=K
\end{eqnarray}
where periodic boundary conditions are implicit. At this stage,
Eqs~\eqref{eq:mean-field}, \eqref{eq:transcurrNTASEPS},
and~\eqref{trcurr} completely define the hydrodynamic mean-field
description of our model.

We will proceed as follows: We first consider left and right
equilibrated reservoirs of equal densities $\rho_1^R=\rho_1^L = \rho_1$; for
each value of $\rho_1$, which is an input parameter, we solve the
equilibrated mean-field equations~\eqref{equilplat} to obtain
$\rho_{i\neq 1}^{L/R}(\rho_1)$, $K(\rho_1)$ and $J_{\rm tot}(\rho_1)$,
which we then  use to predict the phase diagram. Let us first compare these
mean-field predictions with the result of microscopic Monte-Carlo
simulations.

\begin{figure}
  \centering
  \includegraphics[width=.328\textwidth]{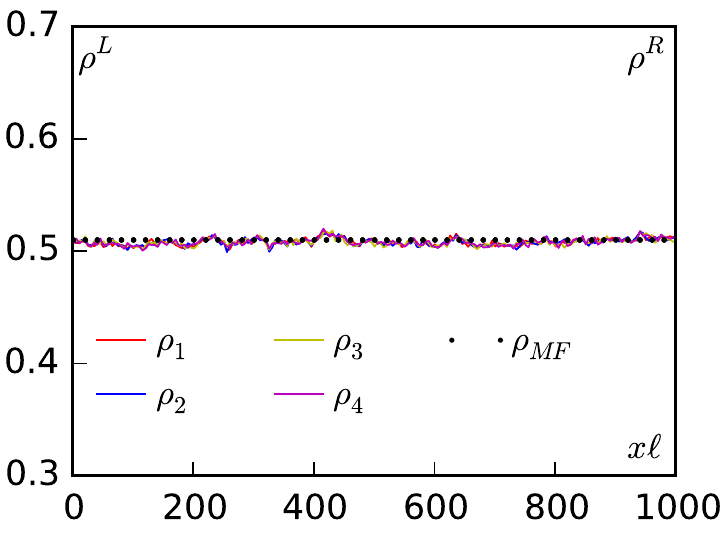}
  \includegraphics[width=.328\textwidth]{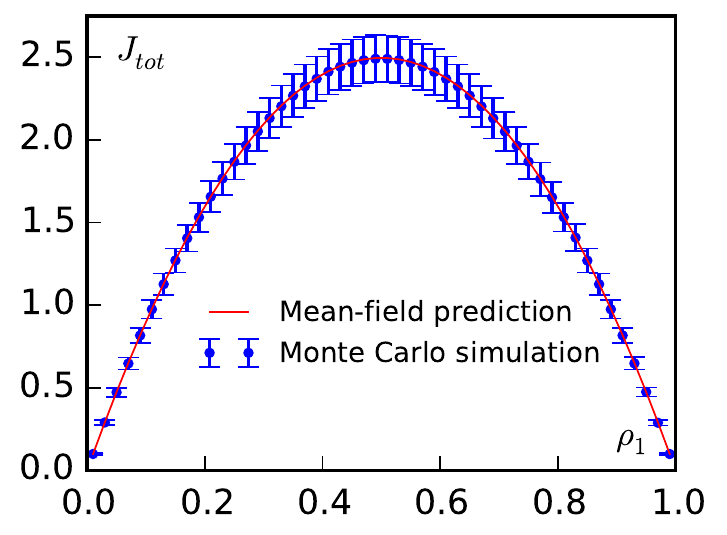}
  \includegraphics[width=.328\textwidth]{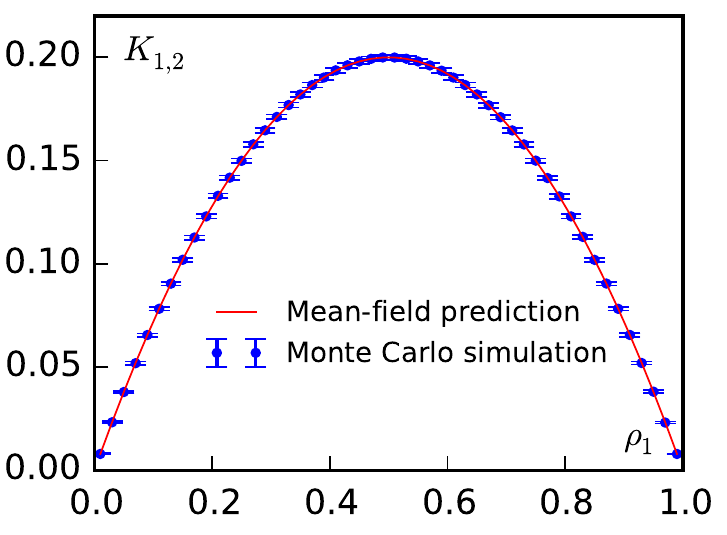}
  \caption{Simulation results for $N=10$ TASEPs on a ring, with
    $p_i=1$ for all lanes and transverse rates: $d_{i,i+1}=0.9$ and
    $d_{i+1,i}=0.1$. {\bf Left:} All density profiles are equal and
    coincide with the mean-field predictions. For sake of clarity,
    only the densities of lanes 1 to 4 are
    shown. ($\rho_1^L=\rho_1^R=0.51$) {\bf Center:} the total
    longitudinal current as a function of $\rho_1$; the maximum is at
    $\rho_1^M=0.5$. {\bf Right:} The transverse current between lane 1
    and 2 corresponds to its mean-field description. As expected, all
    other currents $K_{i,i+1}$ are equal to $K_{1,2}$.}
  \label{fig:JtotTASEP1}
\end{figure}

We show in Fig.~\ref{fig:JtotTASEP1} the results of simulations for
$N=10$ TASEPs on a ring, with $p_i=1$ for all lanes and transverse
rates: $d_{i,i+1}=0.9$ and $d_{i+1,i}=0.1$. For such transverse rates,
the equilibrated plateau condition simply forces the densities of all
lanes to be equal.  As predicted, we observe plateau profiles of
densities $\rho_i(x)\simeq \rho_i^L=\rho_i^R$. The currents
$J(\rho_1)$ and $K(\rho_1)$ are in agreement with their mean-field
predictions. We can now use the generalised extremal current principle
to construct the phase diagram of the system.

\begin{figure}
  \centering
  \includegraphics[width=.4\textwidth]{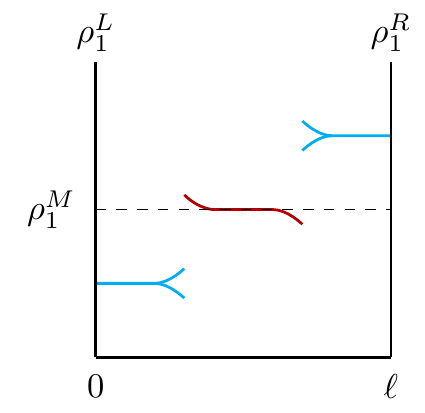}
  \includegraphics[width=.4\textwidth]{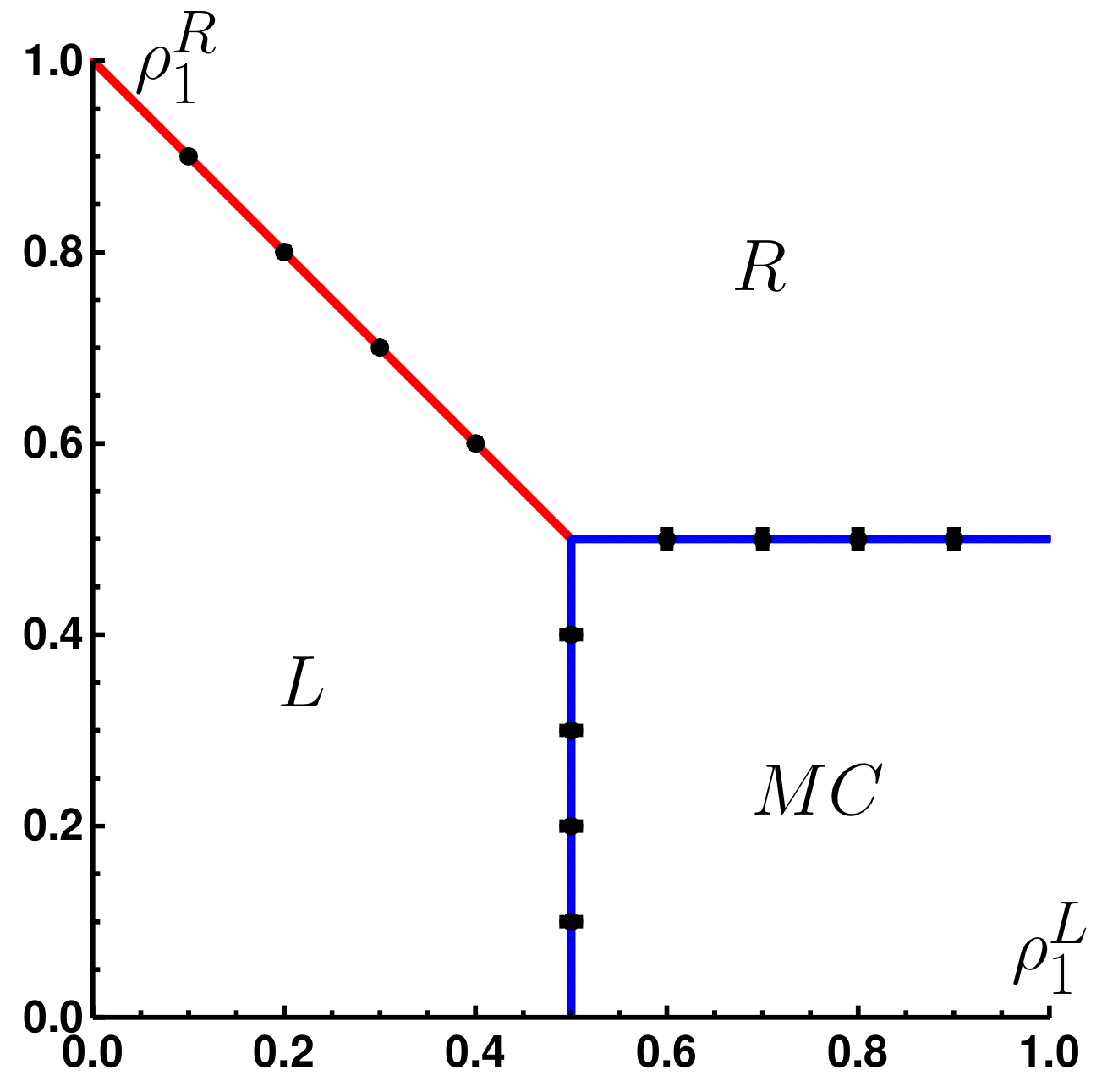}
  \caption{Phase diagram for $N=10$ TASEPs on a ring, with
      $p_i=1$ for all lanes and transverse rates $d_{i,i+1}=0.9$ and
      $d_{i+1,i}=0.1$. {\bf Left:} possible connections of
    equilibrated plateaux predicted by the linear stability
    analysis. {\bf Right:} phase diagram where blue lines indicate
    second order phase transitions and the red line indicates a first
    order phase transition. Black error bars indicate the phase
    boundaries obtained by Monte Carlo simulations. The different
    phases are left (L), right (R) and maximal current phase (MC).}
  \label{fig:phdiagTASEP}
\end{figure}

We proceed as in section~\ref{LSA:phasediag}: we first identify the
extrema of the current and the regions where equilibrated plateaux can
be connected to equilibrated reservoirs at different densities on
their left or right ends (see the left panel of
Fig.~\ref{fig:phdiagTASEP}). We then use the extremal current
principle to construct the phase diagram. As show in the right panel
of Fig.~\ref{fig:phdiagTASEP}, the agreement between the predicted
phase diagram and the results of numerical simulations is very good.

The phase diagram shown in Fig.~\ref{fig:phdiagTASEP} is identical to
that of a single TASEP. This shows that the frequently made assumption
that the motion of molecular motors along a microtubule can be
modelled by a single TASEP is indeed
valid~\cite{KL03,Tailleur2009PRL}. Note, however, that the transverse
current $K$ is generically non-zero along the system ({it follows the
  form of the longitudinal current $J$}).  Non-zero transverse
currents have been observed experimentally, for instance for molecular
motors that have helical trajectories along
microtubules~\cite{helical2,helical3,helical1}. Our analysis thus
covers the collective dynamics of such motors, suggesting that their
propensity to form traffic jams should be identical to that of more
standard motors~\cite{Tailleur2009PRL,LPVHDH2012}.

For the particular choice of rates made in this section, the
equilibrated densities are equal across all lanes. The system thus has
a particle-hole symmetry, and both $J(\rho_1)$ and $K(\rho_1)$ are
symmetric with respect to $\rho_1=0.5$ (see right panels of
Fig.~\ref{fig:JtotTASEP1}). On the first-order transition line (see Fig. \ref{fig:phdiagTASEP}), $K$ is
therefore the same on both sides of a shock. We now turn to a more
general system where this symmetry is violated and show that one can
have coexistence between plateaux with very different values of the
transverse current~$K$.
\begin{figure}
  \centering
  \includegraphics[width=.7\textwidth]{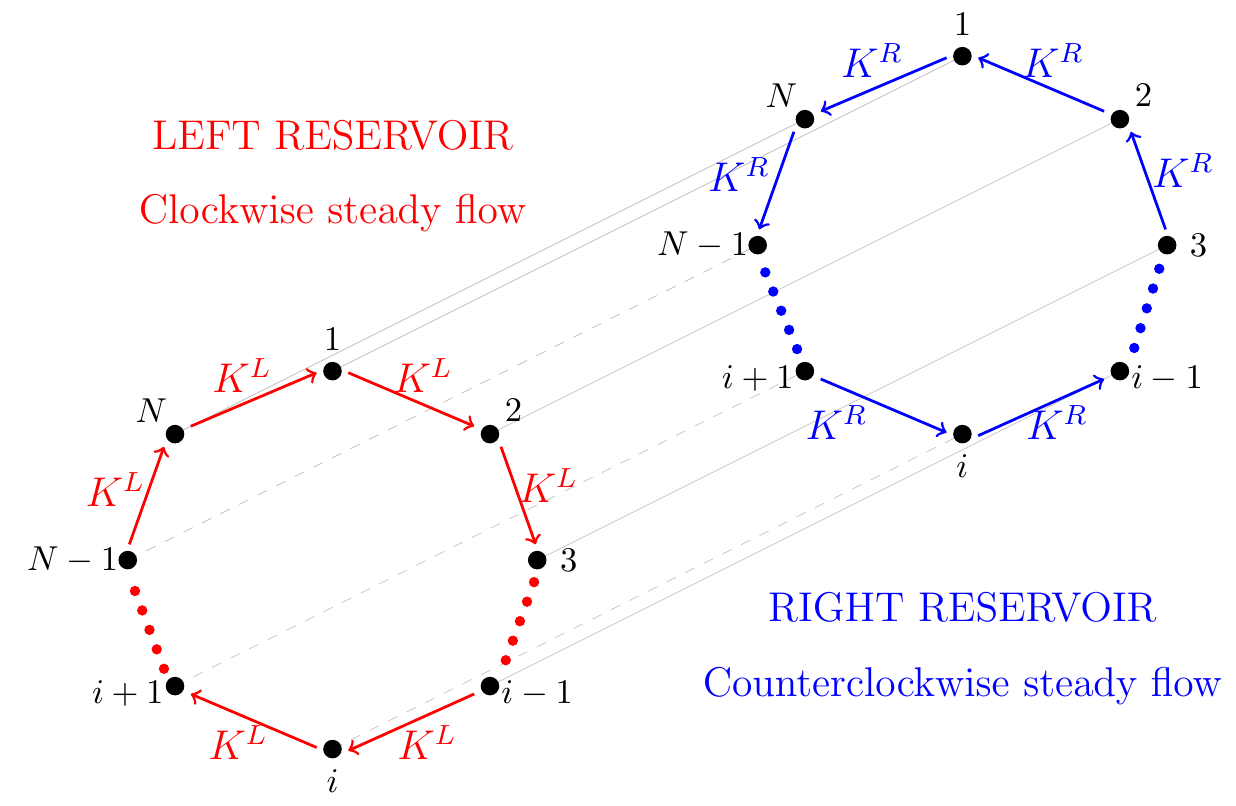}
  \caption{Sketch of a system where $K(\rho^L)$ is positive and
    $K(\rho^R)$ is negative. }
  \label{fig:kloop2}
\end{figure}

\subsection{Counter-rotating shocks}
\label{sec:simuTAPEP}
Let us now turn to more intriguing cases where the transverse current
is not simply proportional to the longitudinal current.
{A particular example we consider is that of a transverse current
  which changes sign along the
lane.
A  way of achieving  a spatial reversal of $K$   would be to have
a transverse current $K(\rho)$  which changes sign as a function of density.}
Specifically, one
could  impose reservoir densities at the two ends of the
system to have, say, clockwise and counter-clockwise flows imposed by
the left and right reservoirs, respectively (see
Fig.~\ref{fig:kloop2}). {Thus, at the boundaries the transverse currents
have opposite signs.}  One can then wonder how the transverse flow
in the bulk   is selected when the system  is ``sheared'' by
such boundary conditions.

In order to achieve a transverse current $K(\rho)$  which changes sign as a function of density, one can first relax the condition that the transverse rates $d_{i,i+1}$ and $d_{i,i-1}$ are equal for all lanes.  As we show in
section~\ref{sec:beyondMF}, the equilibration condition then does not
impose equal densities on all lanes. The transverse current $K$
generically loses its particle-hole symmetry but, as we show
in~\ref{app:transverseK}, it cannot change sign {as a function
  of density} for transverse rates
of the form $d_{i,i\pm1} \rho_i (1-\rho_{i\pm 1})$. 

As we now show, a
partial exclusion process, in which
  the restriction to at most one particle per site is relaxed,
allows us to obtain the desired change of sign in the
transverse current 
  Specifically, we study $N$ parallel one-dimensional
lattice gases in which up to $N_{\rm max}$ particles are allowed on
each site. Particles hop from site $j$ to $j+1$ on lane $i$ with rate
\begin{equation}
  W(i,j\to i,j+1) = p_i \frac{n_{i,j}}{N_{\rm max}} \left ( 1-\frac{n_{i,j+1}}{N_{\rm max}}\right)\;
\end{equation}
%so that $W(i,j\to i,j+1)$ vanishes linearly as the target occupancy
%reaches $n_{i,j+1}=N_{max}$.  
{and we refer to this as the Totally Asymmetric Partial
  Exclusion Process (TAPEP).}
For the transverse hopping rates from
site $j$ of lane $i$ to site $j$ of lane $i\pm 1$, we choose
\begin{equation}
  W(i,j\to i\pm 1,j) = d_{i,i\pm 1} \frac{n_{i,j}}{N_{\rm max}} \left ( 1-\frac{n_{i\pm 1,j}}{N_{\rm max}}\right)
\end{equation}
apart from the hopping from lane 1 to 2, in which we {impose}  a
different hopping rate
\begin{equation}\label{eqn:weirdKmicro}
  W(2,j\to 1,j) = d_{2,1} \left(\frac{n_{2,j}}{N_{\rm max}}\right)^2 \left ( 1-\frac{n_{1,j}}{N_{\rm max}}\right)\;.
\end{equation}
{Note that in the case of the TASEP ($N_{\rm max} =1$)
the hopping dynamics from lane 1 to 2 is the same as that between other lanes since
$n_{2,j}^2 = n_{2j}$. Therefore we need to  consider partial exclusion
($N_{\rm max} >1$) to observe novel behaviour.}

\begin{figure}
  \centering
  \includegraphics[width=.49\textwidth]{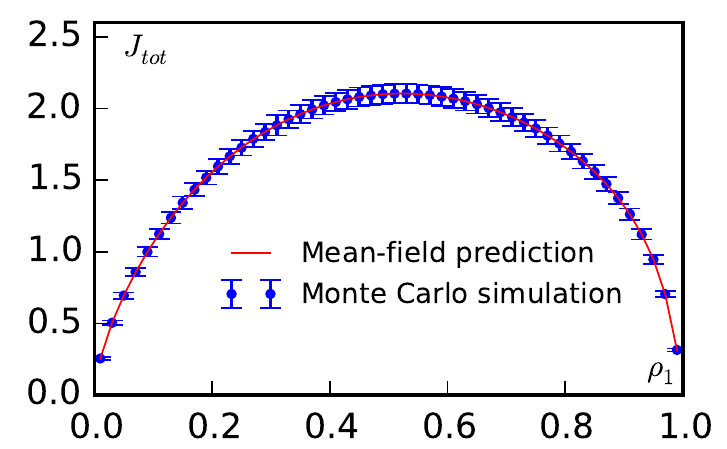}
  \includegraphics[width=.49\textwidth]{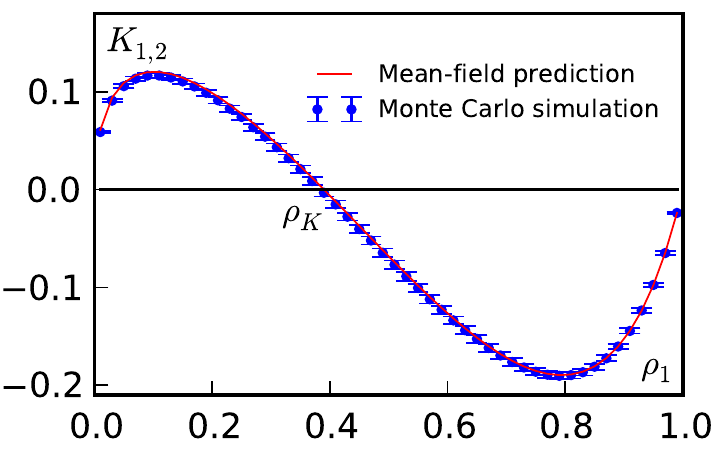}
  \caption{
{Comparison of  numerical
simulations of 10 TAPEPs with mean-field predictions.}
{\bf Left:} total longitudinal current as a function of
    $\rho_1$: the maximum is at $\rho_1^M=0.52$. {\bf Right:}
    transverse current between lane 1 and 2 (currents between all
    other pairs of lanes are the same). In the simulation we have set
    $N_{\text{max}}$=100 and $p=1$ for all the lanes. The transverse
    rates are: $d_{1,2}=56$, $d_{2,3}=89$, $d_{3,4}=7$, $d_{4,5}=27$,
    $d_{5,6}=49$, $d_{6,7}=45$, $d_{7,8}=18$, $d_{8,9}=8$,
    $d_{9,10}=50$, $d_{10,1}=62$, $d_{1,10}=47$, $d_{2,1}=78$,
    $d_{3,2}=95$, $d_{4,3}=94$, $d_{5,4}=10$, $d_{6,5}=41$,
    $d_{7,6}=8$, $d_{8,7}=15$, $d_{9,8}=15$, $d_{10,9}=79$.}
  \label{fig:JtotTAPEP}
\end{figure}
\begin{figure}
  \centering
  \includegraphics[width=.4\textwidth]{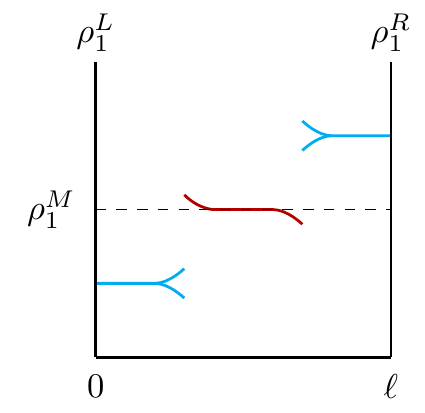}
  \includegraphics[width=.4\textwidth]{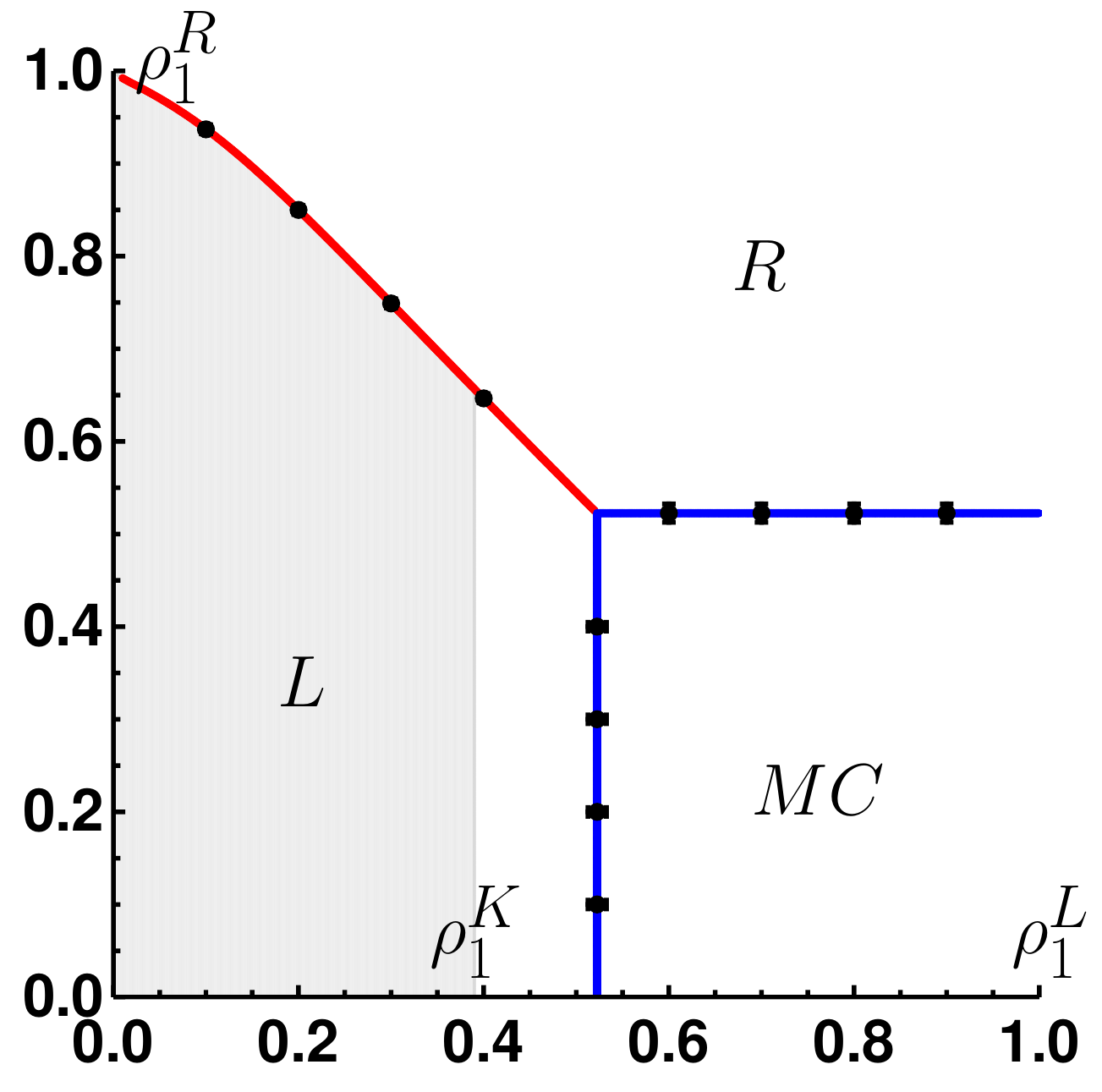}
  \caption{Phase diagram of $N=10$ TAPEPs with $N_{\text{max}}=100$.
    {\bf Left:} possible connections of equilibrated plateaux
    predicted by the linear stability analysis. {\bf Right:} phase
    diagram where blue lines indicate second order phase transitions
    and the red line indicates first order phase transitions. The
    different phases are left, right or maximal current phases. The
    gray shaded region corresponds to the region where $K>0$ is
    observed in the bulk (bottom-left); in the rest of the phase
    diagram, one observes $K<0$ in the bulk. The rates are the same as
    in Fig~\ref{fig:JtotTAPEP}. }
  \label{fig:phdiagTAPEP}
\end{figure}
To derive the hydrodynamic mean-field description of our model, we
proceed as before, to obtain  a dynamics given
by~\eqref{eq:mean-field}, where we have introduced the mean densities
\begin{equation}
  \rho_{i,j}=\frac{\langle n_{i,j}\rangle}{N_{\rm max}}\;.
\end{equation}
Advective and transverse currents are again given
by~\eqref{eq:transcurrNTASEPS} and ~\eqref{trcurr}, apart from
$K_{1,2}$ whose expression is:
\begin{equation}\label{eqn:weirdK}
  K_{1,2} = d_{1,2}\rho_1(1-\rho_{2})- d_{2,1}(\rho_{2})^2(1-\rho_1). 
\end{equation}
Using the methodology introduced in {section~\ref{LSA:phasediag}}, we can now compute,
for equilibrated plateaux, the transverse and longitudinal currents
$K(\rho_1)$ and $J_{\rm tot}(\rho_1)$.

\begin{figure}
  \centering
  \includegraphics[width=.42\textwidth]{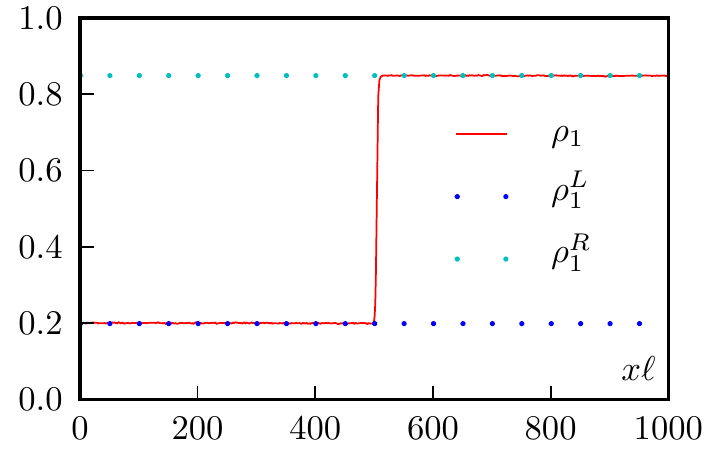}
  \includegraphics[width=.42\textwidth]{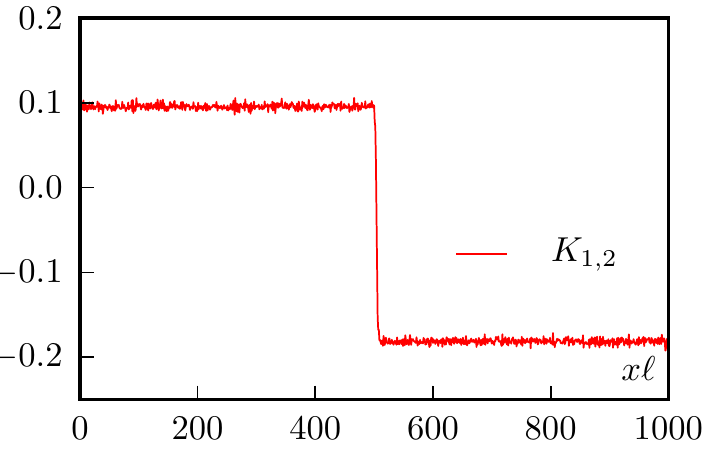}
  \includegraphics[width=.5\textwidth]{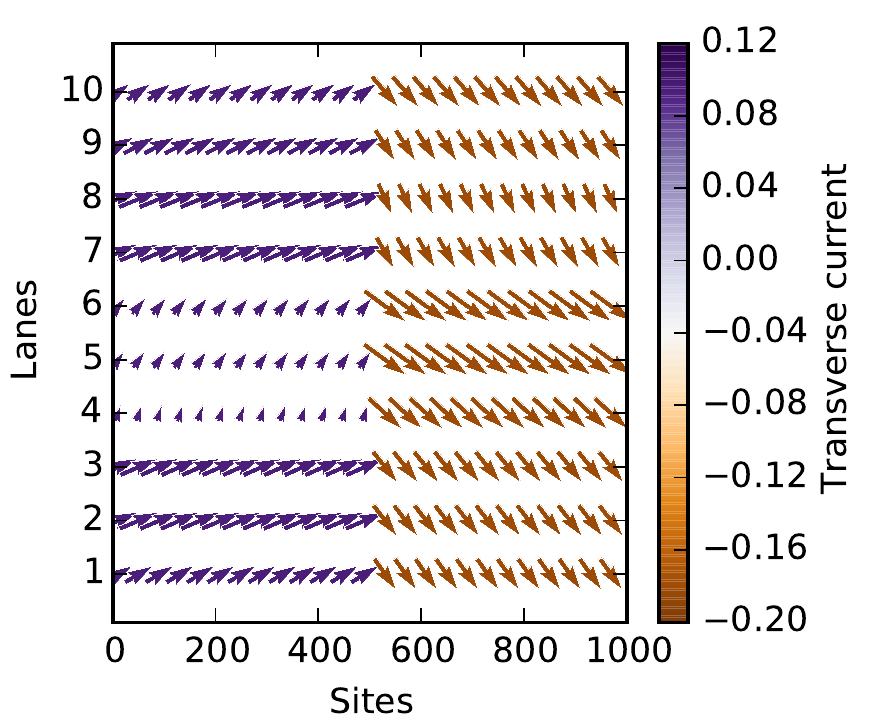}

  \caption{Simulations of $N=10$ TAPEPs with up to
    $N_{\text{max}}=100$ particles allowed per site. All longitudinal
    rates are equal ($p_i=p=1$) whereas the transverse rates are
    chosen as follows: $d_{12}=56$, $d_{21}=47$, $d_{23}=89$,
    $d_{32}=78$, $d_{34}=7$, $d_{43}=95$, $d_{45}=27$, $d_{54}=94$,
    $d_{56}=49$, $d_{65}=10$, $d_{67}=45$, $d_{76}=41$, $d_{78}=18$,
    $d_{87}=8$, $d_{89}=8$, $d_{98}=15$, $d_{9\,10}=50$,
    $d_{10\,9}=15$, $d_{10\,1}=62$, $d_{1\,10}=79$.  {\bf Top-left:}
    the density profile of lane 1 exhibits a shock between a left and
    a right phase.  The densities imposed at the two reservoirs are
    $\rho_1^L=0.199$ and $\rho_1^R=0.849$.  {\bf Top-right:} The
    transverse current between lane 1 and 2 similarly shows a shock
    between two values of opposite signs. {\bf Bottom:} Vector field
    of the current $(J_i,K_{i,i+1})$ along the lattice. The color code
    shows the value of $K$. The length of the arrows corresponds to
    the modulus of the current vector.}
  \label{fig:profilesTAPEP}
\end{figure}

In Fig.~\ref{fig:JtotTAPEP}, we compare the results of numerical
simulations of 10 TAPEPs with mean-field predictions. Again, the
agreement is excellent. {Significantly}, for the particular choice of
transverse rate~\eqref{eqn:weirdKmicro}, $K(\rho_1)$ changes sign as
$\rho_1$ varies, at $\rho_1\equiv \rho_1^K$. Using the expression of
$J_{\rm tot}(\rho_1)$ and the generalised extremal current principle,
we derive the phase diagram which is compared with Monte-Carlo
simulations in Fig.~\ref{fig:phdiagTAPEP}.

On the the red line of Fig.~\ref{fig:phdiagTAPEP}, one observes
coexistence between two profiles of different densities separated by a
shock. {On this line, for $\rho^L<\rho_1^K$}, these shocks separate
the bulk of the system into homogeneous phases with
counter-rotating transverse currents (see
Fig.~\ref{fig:profilesTAPEP}).

\subsection{Shear localisation}
\label{sec:symmcase}
\begin{figure}
  \centering
  \includegraphics[width=.49\textwidth]{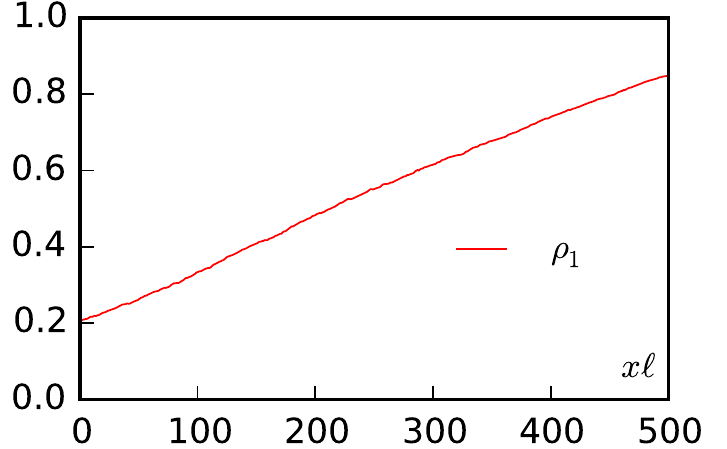}
  \includegraphics[width=.49\textwidth]{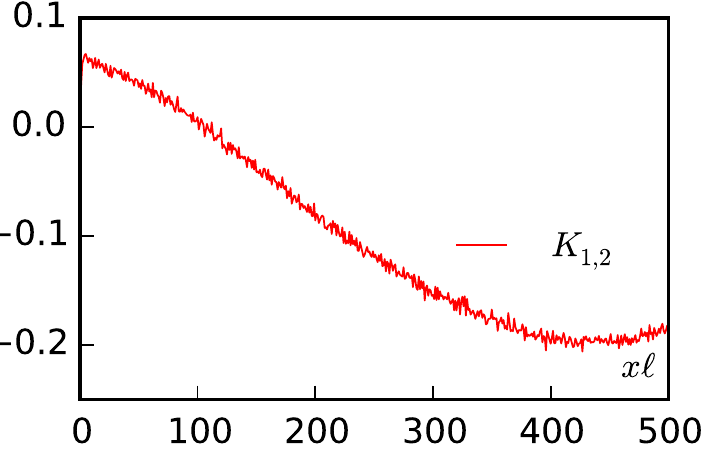}
  \caption{Simulations of 10 Symmetric Partial Exlusion Processes. The
    simulation details are the same as in Fig.~\ref{fig:profilesTAPEP}
    with the exception of the longitudinal dynamics which is now
    symmetric, i.e. $p=q=1$ for all lanes, where $p$ and $q$ are the
    forward and backward hopping parameters respectively. {\bf Left:}
    density profile of lane 1 linearly interpolates between the left
    and right boundary conditions. The density values imposed at the
    two reservoirs are $\rho_1^L=0.199$ and $\rho_1^R=0.849$. {\bf
      Right:} transverse current between lane 1 and 2 continuously
    interpolates between the left and right boundary conditions.}
  \label{fig:SymmCase}
\end{figure}

The phenomenology shown in Fig.~\ref{fig:profilesTAPEP} is rather
counterintuitive. {Indeed, for a Symmetric Partial Exclusion Process
  which yields diffusive dynamics, one expects the density profile to
  linearly interpolate between the two reservoir densities. This in
  turn leads to a continuous variation of the transverse current $K$
  (See Fig.~\ref{fig:SymmCase}).  In contrast, for the driven-diffusive
  dynamics of the TAPEP, which leads to the formation of shocks, our
  analysis shows that a localized discontinuity of the transverse
  current may occur.  Thus in the type of driven system we consider,
  it is the driven longitudinal dynamics which determine the phase
  behaviour and the transverse currents are dictated by the 
  longitudinal ones.

  For standard fluids, where momentum is conserved, such a shear
  discontinuity is unexpected~\cite{Larson}. Multilane models, based
  on random particle hopping, however describe situations where the
  momentum of the system is not even locally conserved. This is
  relevant, for instance, to models of molecular motors hopping along
  a microtubule without a proper description of the surrounding
  fluid. The shear localization phenomenon thus belongs to the class
  of surprising phenomena which can be observed in momemtum
  non-conserving systems (see~\cite{pressure} for how the concept of
  pressure can fail for such systems).}

\subsection{Beyond mean-field theory}

\label{sec:beyondMF}

The approach presented in this paper relies on mean-field
hydrodynamic descriptions of lattice-gases.  We employed a simple
mean-field approximation involving the factorisation of all density
correlations.  In the aforementioned examples, this approach seems to
perfectly predict the phase diagram observed in Monte-Carlo
simulations. As we now show, this is not always the case and the
simple mean-field prediction for $J_{\rm tot}(\rho_1)$ may
fail. Despite of this, the extremal current principle can still be
applied, but with the \textit{numerically measured} current-density
relation $J(\rho_1)$. This suggests that our approach solely relies on
the existence of a hydrodynamic equation, Eq.~\eqref{mean-field}, and not on
the particular mean-field procedure we used here to derive it. Note
that the same is true for single lane systems. In~\cite{Hager} an exact current-density relation, differing from the mean-field prediction, was used to
construct the phase diagram of a single-lane interacting lattice gas
using the extremal current principle.

\begin{figure}
  \centering
  \includegraphics[width=.328\textwidth]{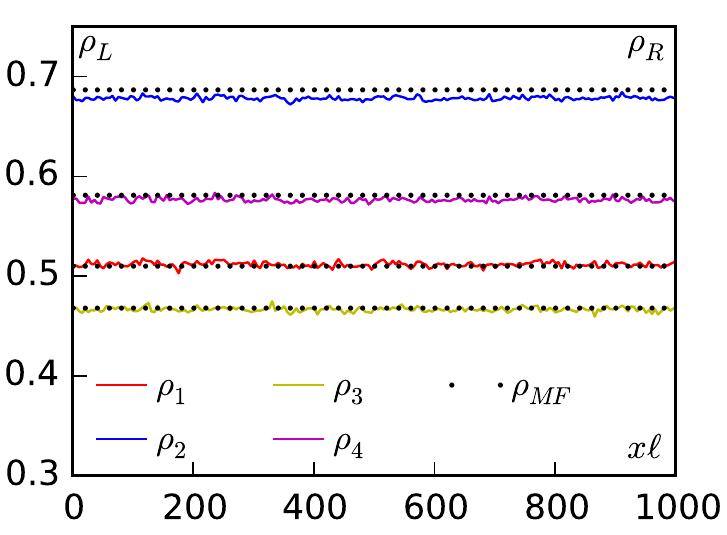}
  \includegraphics[width=.328\textwidth]{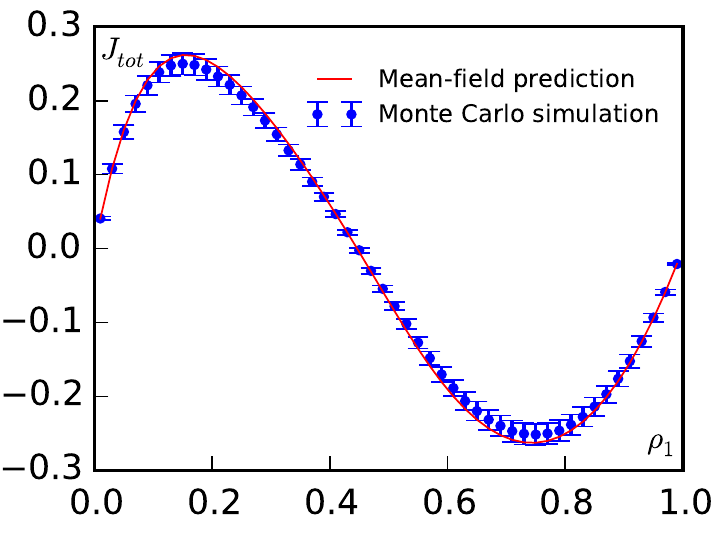}
  \includegraphics[width=.328\textwidth]{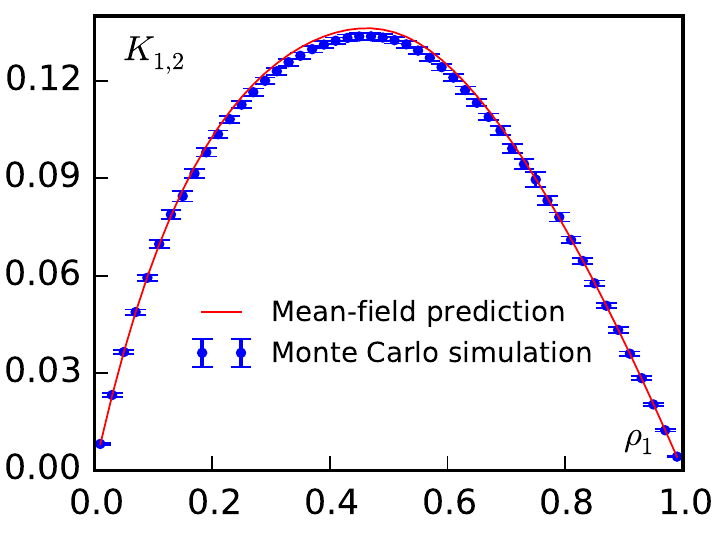}
  \caption{Simulation results for $N=10$ TASEPs on a ring. Even lanes
    ($i=2,4,6,8,10$) correspond to right-going TASEPs whereas odd lanes
    ($i=1,3,5,7,9$) correspond to left-going TASEPs; in all cases, the
    hopping rate is $p_i=1$. The transverse rates are: $d_{1,2}=0.95$,
    $d_{2,3}=0.55$, $d_{3,4}=0.72$, $d_{4,5}=0.62$, $d_{5,6}=0.79$,
    $d_{6,7}=0.99$, $d_{7,8}=0.81$, $d_{8,9}=0.61$, $d_{9,10}=0.93$,
    $d_{10,1}=0.53$, $d_{2,1}=0.05$, $d_{3,2}=0.45$, $d_{4,3}=0.02$,
    $d_{5,4}=0.32$, $d_{6,5}=0.19$, $d_{7,6}=0.49$, $d_{8,7}=0.01$,
    $d_{9,8}=0.41$, $d_{10,9}=0.03$, $d_{1,10}=0.23$. {\bf Left:}
    profiles of densities of lanes 1 to 4 are shown for
    $\rho_1^L=\rho_1^R=0.51$.  {\bf Center:} total longitudinal
    current as a function of $\rho_1$: the maximum is at
    $\rho_1^M=0.15$ while the minimum is at $\rho_1^m=0.75$ (Monte
    Carlo values). {\bf Right:} transverse current between lane 1 and
    2 (currents between all other pairs of lanes are the same).}
  \label{fig:JtotTASEP2}
\end{figure}

Let us now illustrate this by considering $N$ parallel TASEPs where
the transverse hopping rates are non-uniform. Furthermore, to allow
for a richer phase diagram than those shown above, we alternate lanes
where the particles hop rightwards, from site $j$ to site $j+1$, with
lanes where they hop leftwards, from site $j$ to site $j-1$. The
non-uniformity of the transverse hopping rates induces correlations
between different lanes which are neglected within mean-field
theory. As we show in Fig.~\ref{fig:JtotTASEP2}, for a given example
with $10$ lanes and hopping rates specified in the caption, the
profiles and currents calculated using the simple mean-field
approximation do not match exactly the ones measured in the
Monte-Carlo simulations.

\begin{figure}
  \centering
  \includegraphics[width=.328\textwidth]{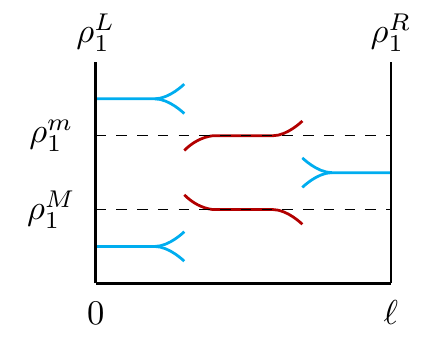}
  \includegraphics[width=.328\textwidth]{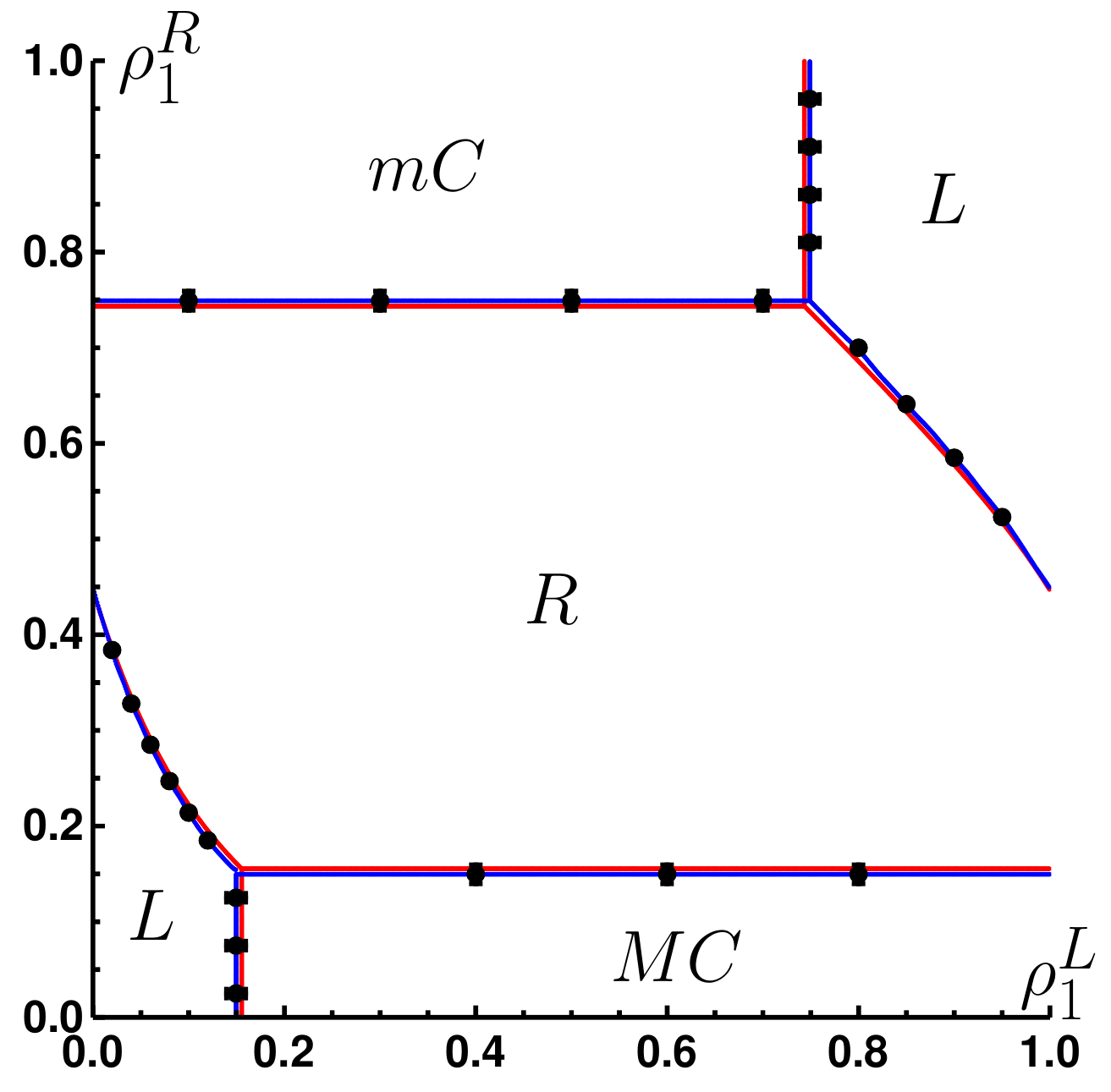}
  \includegraphics[width=.328\textwidth]{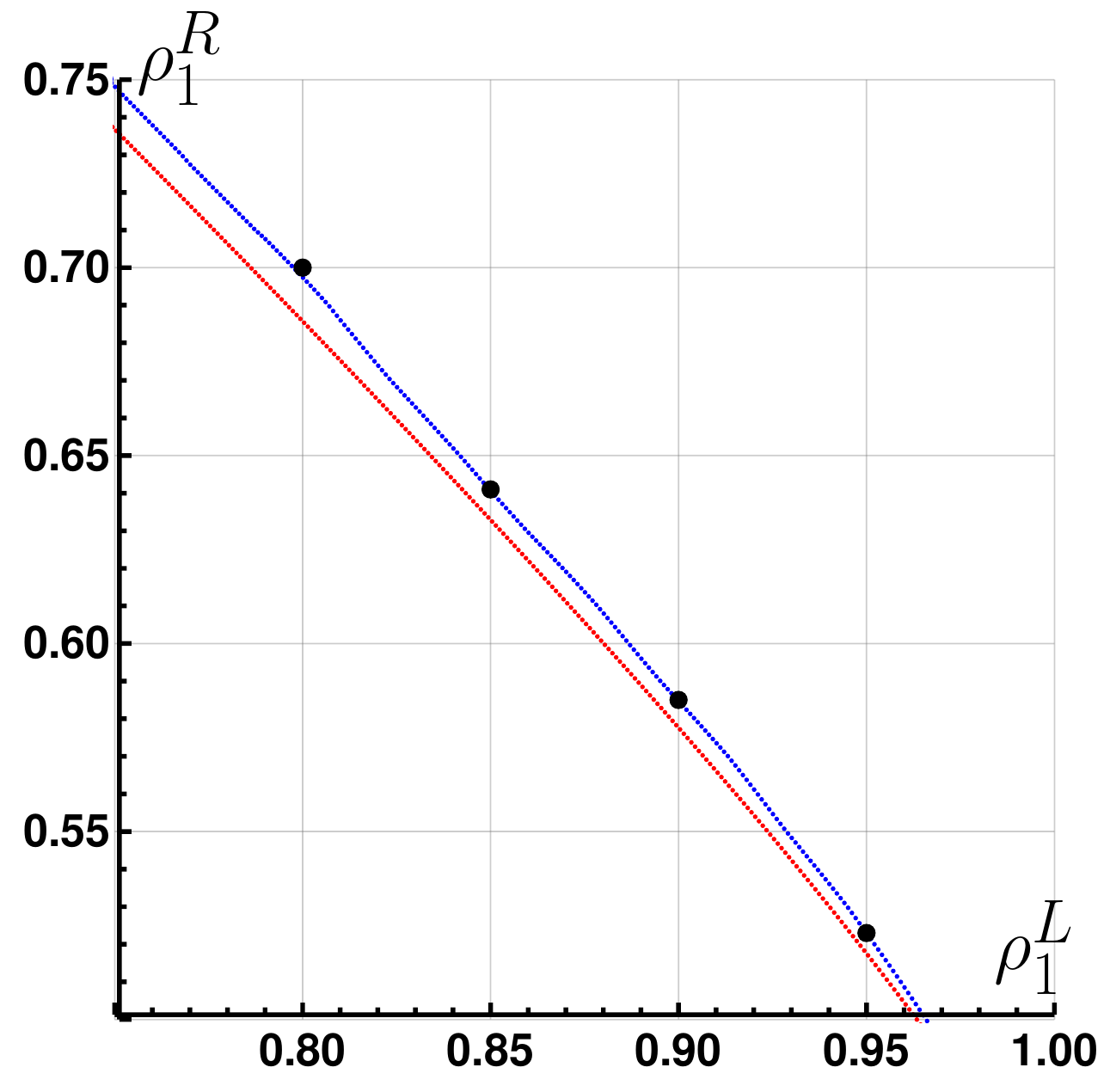}
  \caption{Simulation results for $N=10$ TASEPs on a ring, with the
    same parameters as in Fig~\ref{fig:JtotTASEP2}. {\bf Left:}
    possible connections of equilibrated plateaux predicted by the
    linear stability analysis. {\bf Center:} phase diagram where now
    red lines represent the result of the extremal current principle
    applied to the mean-field prediction and blue lines represent the
    result of the extremal current principle applied to the Monte
    Carlo simulation of $J_{\rm tot}$. Black dots represent numerical
    simulations (with quite invisible errorbars) in the vicinity of
    phase transition lines. The possible phases are Left Phase, Right
    Phase, Maximal Current Phase and minimal Current Phase. {\bf
      Right:} zoom on a region where the discrepancy between
    mean-field prediction and Monte Carlo simulation results are
    appreciable.}
  \label{fig:phdiagTASEPmfpaster}
\end{figure}

Consequently, the phase-diagram predicted by the mean-field expression,
$J_{\rm tot}(\rho_1)$, is slightly off its Monte-Carlo
counterpart. However, applying the extremal current principle instead on
the numerically measured current-density relation yields a phase
diagram which is in perfect agreement with the Monte-Carlo simulations (see
Fig.~\ref{fig:phdiagTASEPmfpaster}).

\section{Conclusions}

In this article we have shown that a generalized extremal current
principle can be used to construct the phase diagram of multilane
driven diffusive systems under the hypothesis that the hopping rate
along one lane does not depend on the occupancies of the neighbouring
lanes, and that the hopping rate between lanes increases with the
occupancy of the departing site and decreases with the occupancy of
the target site. This allowed us to show that the phase diagrams of
such systems are equivalent to those of single-lane systems, though
with much more complicated current-density relations. It validates the
frequently made hypothesis that molecular motors hopping along
microtubules can be effectively describe by single lane models. In all
our modelling, we have used equilibrated reservoirs; all our results
extends to the case of non-equilibrated reservoirs but with two new
boundary layers connecting the bulk equilibrated plateaux with the
reservoirs, as shown in Fig~\ref{fig:unequilibratedreservoirs}. In
this case, the relation between the densities of the bulk plateaux and
those of the reservoirs are, however, not known in general.

Our theory is based on the analysis of hydrodynamic descriptions of
lattice gases. To apply our approach to precise microscopic models one
thus needs to construct such continuous descriptions, for instance
using mean-field approximations such as the simple mean-field approximation
we have employed here. Those are known to fail when
correlations between neighbouring sites are important but we have
shown that the generalized extremal current principle still holds with
respect to the exact current-density relation which can be measured in
a simulation.

The existence of transverse currents, impossible for single and two
lane systems, nevertheless leads to interesting phenomenology. For
instance, we have shown that when a system has different transverse
flows imposed at its boundaries, driven-diffusive systems can have a
bulk behaviour strongly different from diffusive systems. For the
latter, the transverse flow smoothly interpolates between its imposed
bundary conditions whereas driven-dffusive systems can exhibit
`shear-localization'. The system then splits into homogeneous parts with
constant transverse flows separated by sharp interface(s).

Our study extends the physics of one-dimensional boundary driven phase
transitions to more complex systems and in particular to 2D
lattices. It would be interesting to further pursue this approach with
more general lattice structures, such as the network of filaments
encountered in active gels~\cite{NKP11}.

\vskip 2em\appendix{}

\section{Dynamical stability of equilibrated plateaux}
\label{app:dynstab}
Consider a small perturbation $\delta \rho$ around the equilibrated
plateau solution (see Section~\ref{sec:EPdef}) which may be decomposed as
\begin{equation}\label{FEpertapp}
  \rho_i(x,t)=\rho_i^0+\sum_q\delta\rho_i^q(t)\exp(iqx)\;,
\end{equation}
where $q=2\pi n$ with $n=1,\ldots,\ell-1$.
We will denote by $|\delta \rho^q)$ the vector 
$(\delta\rho_1^q, \ldots \delta \rho_N^q)$.
Inserting expression (\ref{FEpertapp}) into the mean-field equations (\ref{mean-field})
and expanding to first order in $\delta\rho_i^q$ yields
the equation
\begin{equation}\label{enq:DSlin}
  \frac{\text{d}}{\text{d}t}\vert\delta\rho^q)=C^q\cdot\vert\delta\rho^q),
\end{equation}
where the matrix $C^q$ is defined by
\begin{eqnarray}
  C_{ii}^q  =  -D_iq^2-\text{i}  -{\rm i}qJ_i^i+\sum_{j\neq i}K_{ji}^i\;; \qquad
  C^q_{ij\neq i}  =  K_{ji}^j,
  \label{dynmtrx}
\end{eqnarray}
with $J_i^i\equiv \partial_{\rho_i}J_i(\rho^0)$. Note that the matrix
$C_{ii}^q$ can be defined for any positive real number $q$ and not
only for the discrete values $q=2\pi n$. As we now show, the eigenvalues
$\lambda_i(q)$ of the matrix $C^q$ always have negative real parts and
$|\delta\rho^q(t))$ thus vanishes at large time.

For $q=0$, $C^q$ is a Markov matrix and none of its eigenvalues
$\lambda_i(0)$ has a positive real part: Eq.~(\ref{dynmtrx}) indeed
shows the sum of each column elements to vanish for $q=0$. Conversely,
when $q\rightarrow\infty$, $\lambda_i(q)\sim-D_iq^2$ and $\lambda_i(q)$
thus has  a negative real part. Physically, the short wave-length perturbations are
stabilized by the diffusive terms while the large wave-length perturbations are
stabilized by the exchange between lanes.
% for whom equilibrated
%plateaux are stable fixed point due to~\eqref{equilplat} and
%\eqref{fluxes}.

The eigenvalues of a matrix are continuous functions in $\mathbb{C}$
of its coefficients. For $C(q)$ to have an eigenvalue with a positive
real part, we need $q^*\in[0,\infty[$ such that at least one
eigenvalue satisfies $\Re(\lambda_i(q^*))=0$, which means
\begin{equation}
 \exists |v)\neq 0\qquad\text{such that}\qquad C^q\vert v)=i\varphi\vert v ), \quad \text{where} \quad\varphi\in\mathbb{R}.
\end{equation}
The matrix $A\equiv C^q-i\varphi\mathbb{I}$ is thus singular. It is
however easy to see that $A^t$, the transpose of $A$, is a strictly
diagonally dominant matrix, i.e.
\begin{eqnarray}
  \vert A^t_{ii}\vert & = & \vert D_iq^2+\Big(\sum_{j\neq
    i}K_{ij}^i\Big)+\text{i}(qJ_i^i+\varphi)\vert \nonumber \\ 
  & > & \sum_{j\neq i}\vert A^t_{ij}\vert=\sum_{j\neq i}\vert
  K_{ij}^i\vert=\vert\sum_{j\neq i}K_{ij}^i\vert \nonumber \;.
\end{eqnarray}
The Gershgorin circle theorem states that all eigenvalues of a square
matrix $B$ can be found within one of the circles centered on $B_{ii}$
of radii $\sum_{j\neq i}\vert B_{ij}\vert$. This implies that strictly
diagonally dominant matrices cannot be singular: $A^t$ and $A$ are
thus invertible. Therefore, the matrix $C(q)$ cannot have purely
imaginary eigenvalues: all its eigenvalues thus have a negative real
part for all $q$. Equilibrated plateaux are thus always dynamically
stable when $J_i$ is only a function of $\rho_i$ and when
equations~(\ref{fluxes}) are satisfied. This can break down for more
general systems, when interactions between the lanes are allowed
for~\cite{EKST11}. 
%(\blue{The fact that strictly diagonally
%  dominant matrix are invertible is called Hadamard Lemma. Should we
%  just cite this or explain a bit more, like in the previous
%  paragraph.})

\section{Connecting bulk plateaux to equilibrated reservoirs using the
  eigenvectors of $M$}

\label{sec:appspectrum}

This appendix is divided in two parts. In the first part, we show that
the spectrum of $M$ is composed of either $(i)$ $N$ eigenvalues with
positive real parts and $N-1$ eigenvalues with negative real parts, or
$(ii)$ $N-1$ eigenvalues with positive real parts and $N$ eigenvalues
with negative real parts.

The second part then shows that these two cases respectively
correspond to plateaux which can only be observed in Left and Right
phases.

\subsection{}
\label{sec:appspectrum1}
We begin by considering a semi-infinite problem with a
reservoir at the left end of the system. Let us assume that the
equilibrated plateaux densities $\rho^{\, \rm p}_j$ are determined by
single-valued (a priori unknown) functions of the reservoir densities
$|\rho^{\, L})$:
\begin{equation}
  \forall j \qquad \rho_j^{\, \rm p}=H_j(\rho_1^{\,L},\dots,\rho_N^{\,L}),
\end{equation}
When the reservoir is equilibrated, one simply has $\rho_j^{\, \rm
  p}=H_j(\rho_j^{\,L})=\rho^{\,L}_j$. We now focus on perturbations
of equilibrated reservoirs and show that the space of perturbations
$|\delta \rho^{\,L})$ that leave $|\rho^{\, \rm p})$ invariant is of
dimension $N-1$.

To see this, let us look at the consequence of an infinitesimal
perturbation $|\delta\rho^{\, L})$ of the reservoir densities. This
perturbation results in a change of bulk equilibrated densities from
$|\rho^{\, \rm p})$ to $|\rho^{\, \rm p})+|\delta \rho^{\, \rm p})$,
with
\begin{equation}
\delta \rho^{\, \rm p}_j=\sum_i \left(\partial_{\rho_i^{\, L}}H_j\right) \delta\rho_i^{\, L}\;.
\end{equation}
which  can be rewritten in  matrix form as
\begin{equation}
|\delta \rho^{\, \rm p}) =   H' |\delta \rho^{\, L})\quad\text{where}\quad H'_{ji}\equiv \partial_{\rho_i^{\, L}}H_j.
\label{Hdef}
\end{equation}
By definition the bulk densities $|\rho^{\,\rm p})+|\delta \rho^{\,\rm
  p})$ are still equilibrated. Therefore, as shown in
section~\ref{sec:lambda0}, $|\delta \rho^{\, \rm p}) = \epsilon
|\delta \rho^{\, \rm 0})$, where $\epsilon$ is a small parameter and
thus
\begin{equation}
   H' |\delta \rho^{\, L}) =
\epsilon |\rho^{\, \rm 0})
\label{Hproj}
\end{equation}
  
%Note that since $H$ depends on $\rho^{r}$, so does the matrix
%$H'$. 
The key observation is that since equation (\ref{Hproj}) holds for any
perturbation $|\delta \rho^{\, L})$, the matrix $H'$ projects any
vector $|\delta \rho^{\, L})$ onto the direction $|\delta\rho^{\, \rm
  0})$. Furthermore, we now show that $H'$ is indeed a projector
satisfying $(H')^2= H'$.
Since  $H'  |\delta\rho^{\, L})\propto |\delta\rho^0)$, the image
 of $H'$ is  $\text{Span}|\delta\rho^0)$.
If we consider an equilibrated  boundary density vector
 $|\rho^{\, L})$
then   $|\rho^{\, L}+\eps \delta \rho^0)$ is also equilibrated, which implies
 $H(|\rho^{\, L}+\eps \delta \rho^0))=|\rho^{\, L})+\eps|\delta
 \rho^0)$. Then by linearisation, $H' |\delta \rho^0)=|\delta
 \rho^0)$. Since (\ref{Hproj}) holds for all $|\delta \rho^{\, L})$,  $H'$ thus satisfies $(H')^2=H'$.

 We may now invoke a basic theorem of linear algebra that for a
 projector, here $H'$, the direct sum of the image space and the
 kernel gives the full $N$ dimensional vector space :
\begin{equation}\label{Rnsplit}
  \mathbb{R}^N=\text{Span}(|\delta\rho^0)) \oplus \ker(H')\;.
\end{equation}
Note that $\ker(H')$ is the space of perturbations $|\delta \rho^{\,
  L})$ that leave $|\rho^{\, \rm p})$ invariant (since
$H'|\delta\rho^{\, \rm r})=|\delta \rho^{\, \rm p})=0$). Therefore the
space of perturbations $|\delta\rho^{\, L})$ that leave the bulk density
vector invariant is of dimension $N-1$.

To connect these perturbations to the eigenvectors of the matrix $M$,
let us consider a given equilibrated plateau $|\rho^{\, \rm p})$ and
ask which reservoir density vectors $|\rho^{\, \rm p}+\delta\rho^{\,
  \rm r})$ it can be connected to through a perturbation
$|\delta\rho(x))$. Such a perturbation can be decomposed using the
eigenvectors of $M$ as in
equation~\eqref{eqn:EVdecomp}:
\begin{equation}
  \vert\delta\rho(x))=\sum_{k=1}^{2N-1}\alpha_k\vert\delta\rho^k) e^{\lambda_k x}\;.
  \label{0sbatta2}
\end{equation}
By definition, $|\delta \rho(x))$ vanishes in the bulk of the system
(see Fig.~\ref{fig:lambdas2}) so there is no $\alpha_0$ term in the
sum.

The sum in \eref{0sbatta2} is restricted to $\Re(\lambda_k)<0$ when
connecting to left reservoirs as terms coming from eigenvectors with
eigenvalues with $\Re(\lambda_k)\geq 0$ would not die out in the bulk.
Since $|\delta \rho(x=0))=|\delta \rho^{\, \rm r})$ , the
perturbations \eqref{0sbatta2} need to span the sets of left reservoir
perturbations that leave $|\rho^{\, \rm p})$ invariant. Since the
vector spaces of such perturbations are $N-1$ dimensional, one needs
at least $N-1$ eigenvectors with $\Re(\lambda_k)<0$.

Conversely, the same reasoning for a semi-infinite system connected to
a reservoir on its right would lead to the conclusion that one needs
at least $N-1$ eigenvectors with $\Re(\lambda_k)>0$.

To conclude the first part of this {appendix}, let us summarize what
we know on the spectrum of the matrix $M$. Since it has $2N$
eigenvectors, one of which is associated with the eigenvalue
$\lambda_0=0$, among the remaining $2N-1$ there must be either $N-1$
eigenvectors with $\Re(\lambda_k)<0$ and $N$ eigenvectors with
$\Re(\lambda_k)>0$ or $N-1$ eigenvectors with $\Re(\lambda_k)>0$ and
$N$ eigenvectors with $\Re(\lambda_k)<0$. 

\subsection{}
\label{sec:appspectrum2}
We now show that the first case, with $N-1$ eigenvectors with
$\Re(\lambda_k)<0$ and $N$ eigenvectors with $\Re(\lambda_k)>0$,
corresponds to a Left Phase.

\begin{enumerate}
\item {\bf Connection to left reservoirs}. As we have shown, 
   there is a space  of
  perturbations $|\delta \rho^{\, \rm r})$ of dimension $N-1$ that leave $|\rho^{\, \rm p})$ invariant. Since
  these are spanned by the $N-1$ eigenvectors with $\Re(\lambda_k)<0$,
  the latter constitute a basis of $\ker [ H'(\rho^{\, \rm p})]$ and one can
  construct a perturbation $|\delta \rho(x))$ that connects $\rho^{\, \rm p}$
  to the reservoir $\rho^{\, \rm p}+\delta\rho^{\, \rm r}$.

  Let us now consider a small equilibrated perturbation
$|\delta
  \rho^{\, \rm r})=\alpha |\delta\rho^0)$
 of $|\rho^{\, \rm r})$
  for which there exists a stationary  profile connecting $|\rho^{\, \rm p})$ to
  $|\rho^{\, \rm r})+\alpha |\delta \rho^0)$.
 Since the profile connects to the
  reservoir at $x=0$, one needs to find a decomposition of
  $|\delta\rho(x))$ such that
    \begin{equation}
      |\delta\rho (0))=\sum_{\Re(\lambda_k)<0}\alpha_{k}|\delta\rho^{k})=\alpha |\delta\rho^0).
      \label{i}
    \end{equation}
    However, since $H'$ is a projection on $|\delta\rho^0)$ and
    $\text{Span}(|\delta \rho^k),\Re(\lambda_k)<0)=\ker(H')$,
    Eq.~\eref{Rnsplit} applies and the intersection between $\ker H'$
    and $\text{Span}[|\delta\rho^0)]$ is the
    empty set. As a consequence, $\delta\rho_i(0)$, which belongs to
    both sets because of~\eqref{i}, is the null vector
    ($\alpha_k=\alpha=0$): the sole equilibrated reservoir to which
    $\rho^{\, \rm p}$ can be connected to the left corresponds to
    $|\rho^{\, \rm r})=|\rho^{\, \rm p})$.

  \item {\bf Connection to right reservoirs}. We now consider only the
    $N$ eigenvectors with $\Re(\lambda_k)>0$. While they still span
    $\ker(H')$, and $|\rho^{\, \rm p})$ can still be connected to
    unequilibrated reservoirs on the right, the $|\delta\rho^k)$ need
    not anymore be in $\ker(H')$. Let us show that $|\rho^{\, \rm p})$ can
    be connected to an equilibrated reservoir on the right with
    $|\delta\rho^{\, \rm r})\neq 0$. Again, this requires a perturbation
    $\vert\delta\rho)$ which satisfies
  \begin{equation}
    \sum_{\Re(\lambda_k)>0}\alpha^{k}\vert\delta\rho^{k})=\epsilon\vert\delta\rho^0),
    \label{label1}
  \end{equation}
  Let us first remember that the eigenvectors $|v^k\rangle$ of
  $M(\rho^{\, \rm p})$ are $2N$-dimensional and that $|\delta\rho^k)$ are
  simply the second half of the components of these vectors. We want to
  show that $|\delta\rho^0)\in
  \text{Span}[|\delta\rho^k),\Re(\lambda_k)>0]$. To do so, we consider
  the vector space $X=\{(x_1,\dots,x_N,0,\dots,0), x_i\in
  \mathbb{R}\}$ of dimension $N$. The eigenvector associated to
  $\lambda=0$ is $\vert
  v^0\rangle=(0,\dots,0,\delta\rho^0_1,\dots,\delta\rho^0_N)$ and
  clearly $\vert v^0\rangle\notin X$. From this we can say that
  \begin{equation*}
    \text{dim}(X+\text{Span}\vert v^0\rangle)=N+1.
  \end{equation*}
  Since one also has $\text{dim}[\text{Span}[\vert
  v^{k}\rangle,\Re(\lambda_k)>0]]=N$ and
  $\text{dim}[(X+\text{Span}\vert v^0\rangle)+(\text{Span}\vert
  v^{k}\rangle)] \le 2N$ (the total space is $2N$ dimensional), it
  follows that
  \begin{eqnarray*}
    \text{dim}[(X+\text{Span}\vert v^0\rangle)+(\text{Span}\vert v^{k}\rangle)] & < & \text{dim}(X+\text{Span}\vert v^0\rangle)+  \\
    &   & +\text{dim}(\text{Span}\vert v^{k}\rangle)=2N+1  \;.
  \end{eqnarray*}
  The intersection between the two spaces is not the empty set
  \begin{equation}
    (X+\text{Span}\vert v^0\rangle)\cap(\text{Span}\vert v^{k}\rangle)\neq\emptyset
    \label{label3}
  \end{equation}
  and there exists a vector $\vert w\rangle\in X+\text{Span}\vert
  v^0\rangle$ of the form $\vert w\rangle=(x_1,\dots,x_N,\epsilon\delta\rho^0_1,\dots,\epsilon\delta\rho^0_N)$
  which is also in $\text{Span}\vert v^{k}\rangle$. This vector can 
  be written as $\vert w\rangle=\sum_{\Re(\lambda_k)>0}\alpha^{k}\vert
  v^{k}\rangle$ and the equality of the second half of the components
  yield
  \begin{equation}
    \forall i\quad\sum_{\Re(\lambda_k)>0}\alpha^{k}\delta\rho_i^{k}=\epsilon\delta\rho_i^0\;.
    \label{label4}
  \end{equation}
  One can thus construct a perturbation, spanned by the $|v_k\rangle$
  with $\Re(\lambda_k)>0$, which is proportional to $|v_k^0\rangle$ and
  connects $|\rho^{\, \rm p})$ to an equilibrated reservoir
  $|\rho^{\, \rm p})+\eps |\delta\rho^0)$ with $\eps\neq 0$.
\end{enumerate}

Conversely, the same reasoning would show that, for equilibrated
reservoirs, a plateau with $N$ eigenvalues with $\Re(\lambda_k)<0$ can
only be observed in a Right Phase.

\section{Transverse current in parallel TASEPs}
\label{app:transverseK}

We will now show that a definition of the transverse current such as
the one given in (\ref{trcurr}) cannot possibly yield a change of sign
in $K$, i.e., $K=0 \Leftrightarrow \rho_i=\rho_{i\pm1}=0,1$.  To show
this, consider at first $N=3$ and let us take $K=0$. Then we have
\begin{eqnarray}
  d_{12}\rho_1(1-\rho_2) & = & d_{21}\rho_2(1-\rho_1) \nonumber
  \\ d_{23}\rho_2(1-\rho_3) & = & d_{32}\rho_3(1-\rho_2) \nonumber
  \\ d_{31}\rho_3(1-\rho_1) & = & d_{13}\rho_1(1-\rho_3).
  \label{trmfK0}
\end{eqnarray}
Solving the system of equation for $\rho_i$ ($i=1,2,3$) yields the
equality $(d_{12}d_{23}d_{31}-d_{21}d_{32}d_{13})(1-\rho_i)\rho_i=0$,
from which we can easily see that either $\rho_i=0$ or $\rho_i=1$,
i.e. the two solutions that we excluded, or
$d_{12}d_{23}d_{31}=d_{21}d_{32}d_{13}$. This last condition
corresponds to an exactly null transverse current for any value of the
density.

This result can be generalized to the case of an $N-$lane system by
using the recursion relation
\begin{equation}
  \rho_{i+1}=\frac{\prod_{j=1}^i d_{j,j+1} \rho_1}{\prod_{j=1}^i
    d_{j,j+1} \rho_1+\prod_{j=1}^i d_{j+1,j}(1-\rho_1)}
\end{equation}
and by imposing the periodic condition at the boundary:
$\rho_{N+1}=\rho_1$. Then we find that in order to have $K=0$ for some
$\rho_i\neq0,1$ one must have $\prod_{j=1}^N d_{j,j+1}=\prod_{j=1}^N
d_{j+1,j}$, which however implies $K=0 \ \forall \ \rho_i\neq0,1$. We
have shown here only the necessary condition, but the sufficent one
can be proved by taking (\ref{trcurr}) and setting $\prod_{j=1}^N
d_{j,j+1}=\prod_{j=1}^N d_{j+1,j}$ for generic $N$.

Note that this appendix shows that $K$ cannot vanish if $\rho_i\neq
0,1$ unless it is always zero. One could imagine, however, that $K$
changes sign discontinuously, without ever satisfying $K=0$. {This
never occurs with the simple choice of rates $K_{ij}$ we have
considered  (which is not
very surprising since $K$ can be shown to be the solution of a
polynomial equation whose coefficients are continuous functions of
$\rho_1$).} More complicated transverse hopping rates, not considered in
this article, leading to multiple possibilities of equilibrated
plateaux for a given value of $\rho_1$, could, however, exhibit more
complicated behaviour.

\end{document}